\documentclass[12pt]{article}
\usepackage{amsfonts,amsmath,amssymb,epsf}
\usepackage{graphics} 
\usepackage{ifpdf}
\ifpdf
\usepackage{epstopdf}
\usepackage{hyperref}
\else
\usepackage[hypertex]{hyperref}
\fi

\advance\textheight by \topskip
\newcounter{thm}

\newcommand\sect[1]{\section{#1}\setcounter{equation}0\setcounter{thm}0} 
                   
\newcommand\void[1]       {}

\newcommand\be            {\begin{equation}}
\newcommand\bea           {\begin{equation}\begin{array}l\displaystyle}
\newcommand\bearll        {\begin{array}{ll}\displaystyle}
\newcommand\ee            {\end{equation}}
\newcommand\eear          {\end{array}}
\newcommand\enl           {\\[1em]\displaystyle}

\newcommand\erf[1] {(\ref{#1})}
\newcommand\labl[1]       {\label{#1}\ee}

\newcommand\arxiv[2]      {\href{http://arXiv.org/abs/#1}{#2}}
\newcommand\doi[2]        {\href{http://dx.doi.org/#1}{#2}}

\newcommand\eps           {\varepsilon}
\newcommand\Fs            {\mathsf{F}}
\newcommand\Ga[1]         {\Gamma\hspace*{-1pt}\big(#1\big)}
\newcommand\Gs            {\mathsf{G}}
\newcommand\Hom           {\mathrm{Hom}}
\newcommand\id            {{\rm id}}

\newcommand\Lb            {\overline L}

\newcommand\one           {{\bf1}}

\newcommand\Rs            {\mathsf{R}}
\newcommand\su            {\mathrm{su}}

\newcommand\Cb            {\mathbb{C}}

\newcommand\Rb            {\mathbb{R}}
\newcommand\Zb            {\mathbb{Z}}

\newcommand\Hc            {\mathcal{H}}
\newcommand\Ic            {\mathcal{I}}

\newcommand\Vc            {\mathcal{V}}

\begin{document}

\thispagestyle{empty}
\def\thefootnote{\fnsymbol{footnote}}
\begin{flushright}
KCL-MTH-07-16\\
0711.0102 [hep-th] 
\end{flushright}
\vskip 5.0em
\begin{center}\LARGE
Perturbed Defects and T-Systems\\
 in Conformal Field Theory
\end{center}
\vskip 4em
\begin{center}\large
  Ingo Runkel\footnote{Email: {\tt ingo.runkel@kcl.ac.uk}}%
\end{center}
\begin{center}
  Department of Mathematics, King's College London \\
  Strand, London WC2R 2LS, United Kingdom  
\end{center}
\vskip 1em
\begin{center}
  November 2007
\end{center}
\vskip 4em
\begin{abstract}
Defect lines in conformal field theory can be perturbed by chiral defect fields. If the unperturbed defects satisfy $\su(2)$-type fusion rules, the operators associated to the perturbed defects are shown to obey functional relations known from the study of integrable models as T-systems. The procedure is illustrated for Virasoro minimal models and for Liouville theory. 
\end{abstract}

\vskip 4em

\begin{center}
  {\em Dedicated to the memory of Alexei Zamolodchikov.} 
\end{center}

\setcounter{footnote}{0}
\def\thefootnote{\arabic{footnote}}

\newpage

\sect{Introduction and summary}\label{sec:intro}

This paper is concerned with purely transmitting defect lines in two-dimensional conformal field theory (CFT), where the defect line itself may break conformal invariance. Examples of such defects can be obtained by perturbing a purely transmitting conformal defect by a chiral defect field \cite{Konik:1997gx,Bachas:2004sy,Fuchs:2007tx}. If the perturbation is relevant, this results in a renormalisation group (RG) flow between two purely transmitting conformal defects. 

Purely transmitting conformal defects are useful because they allow one to deduce symmetries and dualities of the CFT \cite{Frohlich:2004ef,defect}. In particular, the defects act on the set of conformal boundary conditions, and they also provide relations among the RG flows between different boundary conditions \cite{Graham:2003nc}. Conversely, by acting with a perturbed defect on unperturbed conformal boundary conditions, a single defect RG flow induces a whole series of boundary RG flows. In this sense defect flows are `universal RG flows' for boundary conditions \cite{Bachas:2004sy}.

Another observation related to boundary RG flows is that in certain cases (e.g.\ integrable boundary perturbations in minimal models) the disc amplitudes for the perturbed boundary conditions obey a system of functional relations \cite{Bazhanov:1994ft,Dorey:1999cj} -- called fusion hierarchy or T-system -- known from the study of integrable lattice models and integrable continuum field theories, see e.g.\ \cite{Baxter,Bazhanov:1987zu,KlPe,Kuniba:1993cn,Bazhanov:1994ft}. The usefulness of these functional relations lies in the fact that, together with certain assumptions on their analytic properties, they can be solved in terms of a set of integral equations known as the thermodynamic Bethe ansatz \cite{Zamolodchikov:1989cf}, see \cite{Dorey:2007zx} for a review.

In this paper a simple proof is given that -- under certain conditions to be described in detail below -- the perturbed defect operators themselves satisfy the T-system functional relations. This result can be used to explain the behaviour of the perturbed disc amplitudes, but it contains much more information than that since the defect operators act on all bulk states, not just on the ground state. It also offers an alternative point of view on the {\bf T}-operators defined in \cite{Bazhanov:1994ft,Bazhanov:1996dr,Bazhanov:1998dq}, which can be thought of as a `chiral part' of a perturbed defect operator. The arguments leading to the functional relations stay within CFT and do not rely on results from integrable scattering theories. One may therefore hope that this result can further clarify the `ODE/IM correspondence', an interesting link between the conformal limit of two dimensional integrable models and the spectral theory of ordinary differential equation \cite{Dorey:1998pt,Bazhanov:1998wj,Dorey:2007zx}.

\bigskip

Let us look in more detail at the properties of defects and their perturbations. A defect line is a line on the surface on which the CFT is defined where fields can have discontinuities or singularities. Just as for surfaces with boundaries one must specify a boundary condition, a defect is characterised by a `defect condition' or {\em defect type}. Consider the CFT on a cylinder, and denote by $\Hc$ the space of states on a circle. A defect line of some type $a$ wound around the cylinder then gives rise to a linear operator $D_a$ on $\Hc$, called {\em defect operator}. The defect is {\em conformal}, iff $[L_m{-}\Lb_m,D_a] = 0$ for all $m\in\Zb$, and it is called a {\em purely transmitting conformal defect}, or a {\em topological defect}, iff the stronger condition $[L_m,D_a] = [\Lb_m,D_a] = 0$ holds. It is also natural to consider defect lines which are actually interfaces joining two different CFTs, see e.g.\ \cite{Bachas:2001vj,defect,Quella:2006de,Fuchs:2007tx}, but we will not use this here. Note, however, that a conformal boundary condition is a special type of defect, namely a conformal defect that joins a given CFT to the trivial CFT (which has $c\,{=}\,0$ and whose only state is the vacuum).

\begin{figure}[tb]
\begin{center}
i)
  \raisebox{-80pt}{
  \begin{picture}(110,95)
   \put(0,0){\scalebox{.75}{\includegraphics{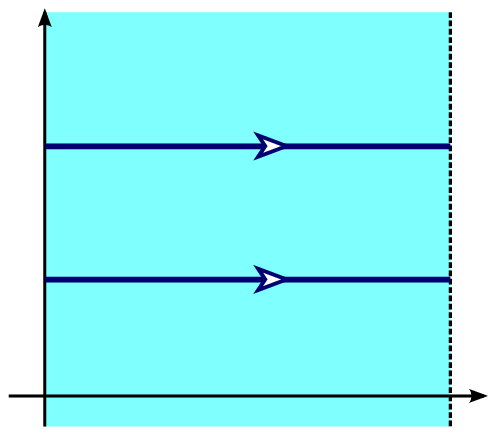}}}
   \put(0,0){
     \setlength{\unitlength}{.75pt}\put(-140,-6){
     \put(142,  4) {\scriptsize$ 0 $}
     \put(272,  4) {\scriptsize$ L $}
     \put(184, 55) {$ b $}
     \put(184, 94) {$ a $}
     \put(138, 46) {\scriptsize$ iy $}
     \put(138, 85) {\scriptsize$ ix $}
     }\setlength{\unitlength}{1pt}}
  \end{picture}}
\hspace*{7em}
ii)
  \raisebox{-80pt}{
  \begin{picture}(110,95)
   \put(0,0){\scalebox{.75}{\includegraphics{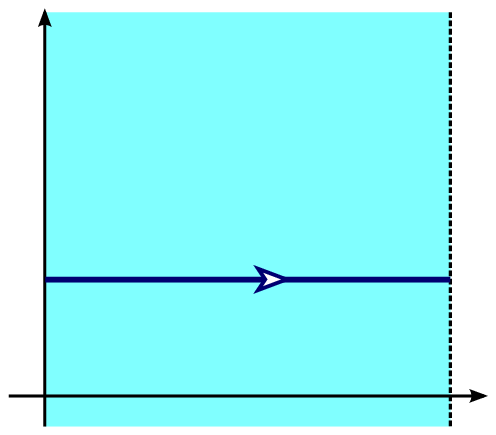}}}
   \put(0,0){
     \setlength{\unitlength}{.75pt}\put(-140,-6){
     \put(142,  4) {\scriptsize$ 0 $}
     \put(272,  4) {\scriptsize$ L $}
     \put(170, 55) {$ a \star b $}
     \put(138, 46) {\scriptsize$ iy $}
     }\setlength{\unitlength}{1pt}}
  \end{picture}}
\end{center}
\caption{In these pictures the lines $i\Rb$ and $i\Rb + L$ are identified. (i) The correlator of the two topological defects $a$, $b$ inserted on the cylinder at $ix$ and $iy$ does not depend on $x$ and $y$. (ii) In the limit $x \rightarrow y$ one obtains the fused defect $a \star b$.}
\label{fig:fused-def}
\end{figure}

The Hamiltonian of a CFT on a cylinder of circumference $L$ is $H(L) = \tfrac{2\pi}L\big(L_0 + \Lb_0 - \tfrac{c}{12}\big)$.
Since topological defects obey in particular $[H(L),D_a] = 0$, correlators do not depend on the precise point at which a topological defect loop is inserted on the cylinder. If two topological defects of type $a$ and type $b$ are inserted on adjacent loops on the cylinder, they can be moved arbitrarily close to each other without encountering a singularity. This results in a new `fused' defect whose type will be denoted by $a \star b$, see figure \ref{fig:fused-def}. Even if the two defects $a$, $b$ we started from are elementary (i.e.\ they cannot be written as a superposition of other defects), the fused defect is typically not, and one has a decomposition $a \star b = c_1 + \cdots + c_n$ in terms of elementary defects $c_k$. This gives rise to the fusion algebra of topological defects \cite{Petkova:2000ip,Petkova:2001ag,Chui:2001kw,Coquereaux:2001di,tft1,defect}. For the defect operators this decomposition implies the identity $D_{a \star b} = D_{c_1} + \cdots + D_{c_n}$.

As already noted in \cite{Konik:1997gx,Bachas:2004sy}, the property $[H(L),D_a] = 0$ -- which was necessary to define the fusion procedure -- continues to hold if we perturb the topological defect by a chiral defect field $\phi(z)$ (i.e.\ a defect field satisfying $\frac{\partial}{\partial\bar z} \phi(z) = 0$). Let us denote the perturbed defect by $D_a(\lambda \phi)$, with $\lambda \in \Cb$ the coupling constant. In section \ref{sec:fun-eq} I present a method to derive functional equations for the operators $D_a(\lambda \phi)$ in a subset of all topological defects where the defects have $\su(2)$-type fusion rules, and for which the perturbing field has conformal weight $< \tfrac12$, so that no regularisation is required. Some examples of models where such subsets exist are non-unitary Virasoro minimal models, Liouville theory, and the $\su(2)$-WZW model for level $k\,{>}\,2$.

Let us consider the minimal model $M(p,p')$ for concreteness. The elementary topological defects are labelled by entries $(r,s)$ in the Kac-table (modulo the usual $\Zb_2$-identification), where $1 {\le} r {<} p$ and $1 {\le} s {<} p'$. The fusion of these defects is just given by the fusion of the corresponding irreducible representations \cite{Petkova:2000ip}. The subset we are interested in consists of the defects labelled by $(1,s)$, and for $s=2,\dots,p'{-}2$ these allow for a chiral defect $\phi$ with weight $h_{1,3}= {-}1{+}2p/p'$. The condition $h_{1,3} < \tfrac12$ thus holds whenever $p/p'<3/4$ (i.e.\ never for unitary models). It is shown in section \ref{sec:ex-mm} that the perturbed defect operators mutually commute,
\be
  \big[\,D_{(r,s)}(\lambda\phi)\,,\,D_{(r',s')}(\mu\phi)\,\big] 
  = 0 
\labl{eq:mm-commute}
for all $(r,s)$ and $(r',s')$ in the Kac-table, and for all $\lambda,\mu \in \Cb$, and that the defect operators labelled $(1,s)$ obey
\be
  D_{(1,2)}(\lambda\phi) \,
  D_{(1,s)}(q^{\eps s} \lambda\phi)
  ~=~
  D_{(1,s-1)}(q^{\eps (s+1)} \lambda\phi)
  \,+\, D_{(1,s+1)}(q^{\eps (s-1)} \lambda\phi)  
\labl{eq:mm-fus-2s}
for $q = e^{\pi i p/p'}$, $s=2,\dots,p'{-}2$, $\eps = \pm 1$ and $\lambda \in \Cb$. In this equation it is understood that for the two defects $(1,1)$ and $(1,p'{-}1)$ which do not support the chiral defect field $\phi$, $D_a(\lambda \phi)$ just stands for the unperturbed defect $D_a$. In particular, $(1,1)$ is the invisible defect, meaning there is actually no defect line present. The defect fields on the $(1,1)$-defect are precisely the bulk fields, and the only bulk fields with anti-holomorphic weight zero are descendents of the vacuum. Also note that $D_{(1,1)}$ is just the identity operator on the space of states $\Hc$. From \erf{eq:mm-fus-2s} it is straightforward to deduce the functional relation of the T-system,
\be
  D_{(1,s)}(q \lambda \phi) \,
  D_{(1,s)}(q^{-1} \lambda \phi)
  ~=~ \id \,+\, 
  D_{(1,s-1)}(\lambda \phi) \,
  D_{(1,s+1)}(\lambda \phi) ~.
\labl{eq:mm-Tsys}
The defect operators $D_{(1,s)}(\lambda \phi)$ thus behave similarly to fused row transfer matrices in RSOS lattice models, which obey functional relations analogous to \erf{eq:mm-fus-2s} and \erf{eq:mm-Tsys} \cite{Bazhanov:1987zu,KlPe}. The $D_{(1,s)}(\lambda \phi)$ are also close cousins of the {\bf T}-operators constructed in \cite{Bazhanov:1994ft,Bazhanov:1996dr,Bazhanov:1998dq}, which equally obey \erf{eq:mm-fus-2s} and \erf{eq:mm-Tsys}.\footnote{
  Equations (4.13) and (4.14) of \cite{Bazhanov:1998dq} 
  are of the form 
  \erf{eq:mm-fus-2s} when setting $\beta^2=p/p'$ 
  (cf.\ (1.27) in \cite{Bazhanov:1998dq}) and replacing 
  $T_j(\lambda) \rightarrow D_{(1,2j+1)}(\lambda^2\phi)$.
  When comparing to 
  \cite{Bazhanov:1994ft,Bazhanov:1996dr,Bazhanov:1998dq}, it should 
  be kept in mind that the {\bf T}-operators are chiral
  operators acting on individual representations, while the 
  $D_{(1,s)}(\lambda \phi)$ act on the whole space of bulk 
  states. The precise relation between the two remains to be
  understood.} 

As reviewed in \cite{Dorey:2007zx}, in order to solve \erf{eq:mm-Tsys} one uses the fact that due to \erf{eq:mm-commute} all $D_{(r,s)}(\lambda \phi)$ can be simultaneously diagonalised, and so the eigenvectors can be chosen to be independent of $\lambda$. The resulting functional equations for the eigenvalues can be turned into integral equations that can be solved very efficiently numerically.

\medskip

It is in fact surprisingly easy to arrive at \erf{eq:mm-fus-2s}. Consider the composition $D_{(1,2)}(\lambda\phi)$ $D_{(1,s)}(\mu\phi)$ for arbitrary $\lambda,\mu \in \Cb$. One first needs to notice that when fusing the perturbed defects $(1,2)$ and $(1,s)$ one obtains the superposition $(1,s{-}1)+(1,s{+}1)$ perturbed by defect fields on $(1,s{-}1)$ and $(1,s{+}1)$, {\em as well as} by defect changing fields which change the defect type from $(1,s{-}1)$ to $(1,s{+}1)$ and vice versa. The defect changing fields stop us from writing the operator for the perturbed defect $(1,s{-}1)+(1,s{+}1)$ as a sum of two defect operators $D_{(1,s-1)}(\lambda' \phi) + D_{(1,s+1)}(\lambda'' \phi)$ for some $\lambda',\lambda''\in\Cb$. However, it turns out to be possible to choose the constant $\mu$ in terms of $\lambda$ such that the defect changing fields are completely suppressed, and only the defect preserving fields contribute to the defect operators. This results in the identity \erf{eq:mm-fus-2s}. On the other hand, for defects $a,b$ which fuse to three or more elementary defects, $a \star b = c_1 + c_2 + c_3 + \cdots $, this construction will typically fail, because there is only one parameter, $\mu$, to adjust, and this is generally not enough to make all the couplings to defect changing fields vanish. (Incidentally, defect changing fields always come in pairs, and so even if $a \star b = c_1+c_2$, there are still {\em two} different defect changing fields. But we will see in section \ref{sec:defch-pert} that it is enough to be able to set one of the two couplings to zero.)

In order to make the above reasoning precise one needs a good control over the OPE of defect fields and over the effect of fusing defects in the presence of defect field insertions. Both are available in the TFT approach to rational CFT \cite{Felder:1999cv,Fuchs:2001am}. Specifically, we will need the results from \cite{tft4,defect}.

\medskip

As an application, let us look at some consequences of \erf{eq:mm-Tsys} on amplitudes involving perturbed boundary conditions. Consider a cylinder of circumference $L$ and length $R$, with conformal boundary conditions at either end labelled by the Kac-label $(1,1)$. Inside the cylinder place two defect loops corresponding to the operators $D_{(1,s)}(q \lambda \phi)$ and $D_{(1,s)}(q^{-1} \lambda \phi)$. One can now fuse each of the two defects with one of the conformal boundaries, resulting in the boundary condition $(1,s)$ perturbed by $q \lambda \phi$ and $q^{-1} \lambda \phi$ (with the appropriate normalisation for the $h_{1,3}$-boundary field $\phi$). This results in the identity
\be
  Z_{(1,s)(1,s)}(q \lambda,q^{-1} \lambda)
  = Z_{(1,1)(1,1)} + 
  Z_{(1,s-1)(1,s+1)}(\lambda,\lambda) 
\ee
for the cylinder partition functions, which has already been observed in \cite{Dorey:2000rv} for the (massive) Lee-Yang model. In the expression $Z_{x,y}(\lambda,\mu)$, $x$ and $y$ refer to the conformal boundary condition, and $\lambda$ and $\mu$ are the coupling constants for the perturbation by $\phi$ on either of the two boundaries. In the $R \rightarrow \infty$ limit each cylinder partition function factors into a product of two disc amplitudes, and one obtains the statement that the perturbed disc amplitudes satisfy the T-system functional relation.

\bigskip

The rest of the paper is organised as follows. In section \ref{sec:fun-eq} the functional equation for defects with $\su(2)$-type fusion rules are derived. In section \ref{sec:ex} minimal models and Liouville theory are treated as examples, and section \ref{sec:conc} contains the conclusions.

\sect{Functional equations for defects operators}
\label{sec:fun-eq}

Fix a rational CFT, and denote by $\Vc$ its chiral algebra. For example, choose a minimal model $M(p,p')$ and let $\Vc$ be the Virasoro (vertex-)algebra, or take the $\su(2)$-WZW model at level $k$ and for $\Vc$ the (vertex algebra constructed from the) affine Lie algebra $\widehat{\su}(2)_k$. Denote by $\Ic$ the finite set indexing the irreducible representations of $\Vc$ and by $\{R_i\,|\,i\,{\in}\,\Ic\}$ the corresponding representations. $R_0$ will refer to the vacuum representation, i.e.\ the representation of $\Vc$ on itself, $R_0 = \Vc$.

I will make the simplifying assumptions that the holomorphic and anti-holomorphic chiral algebra are identical, and that the fusion rule coefficients of the irreducible representations of $\Vc$ are either $0$ or $1$,
\be 
  N_{ij}^{~k} \in \{0,1\} ~.
\ee
Although the methods presented below can be applied to general rational CFTs, in the present paper only the Cardy case is considered. In particular, the space of states on a circle is given by
\be
  \Hc = \bigoplus_{k \in \Ic} R_k \otimes \bar R_{\bar k} ~~.
\labl{eq:H-space}
The notation $\bar R$ indicates that the anti-holomorphic copy $\bar\Vc$ of $\Vc$ acts on this factor of the tensor product, and $\bar k$ labels the representation conjugate to $R_k$ in the sense that it is the unique index for which $N_{k\bar k}^{~~0}=1$.

\subsection{Unperturbed topological defects}

Denote the modes generating the holomorphic copy of $\Vc$ by $W_m$ and those generating the anti-holomorphic copy by ${\overline W}_m$. In this section we will consider only defects that preserve $\Vc \otimes \bar \Vc$, i.e.\ whose defect operators $D_a$ on the cylinder obey
\be
  [W_m,D_a] = 0 = [{\overline W}_m,D_a]
\labl{eq:commute-W}
for all modes $W_m$ and ${\overline W}_m$. Since $\Vc$ contains the Virasoro algebra, such defects are in particular topological. From here on `unperturbed defect' or just `defect' refers to a defect satisfying \erf{eq:commute-W}.

It turns out that just as for boundary conditions \cite{Cardy:1989ir}, in the Cardy case defects are labelled by irreducible representations of $\Vc$ \cite{Petkova:2000ip}. Because of \erf{eq:commute-W} the defect operator $D_a$ for $a \in \Ic$ will act as a multiple of the identity on each sector $R_k \otimes \bar R_{\bar k}$ of the space of states. The coefficients can be given in terms of the modular $S$-matrix, and the resulting fusion rules of the defects are just the fusion rules for the representations of $\Vc$ \cite{Petkova:2001ag,defect},
\be
  D_a \big|_{R_k \otimes \bar R_{\bar k}} = 
  \frac{S_{ak}}{S_{0k}} \,
  \id_{R_k \otimes \bar R_{\bar k}}
  \quad \text{and} \quad
  a \star b = \sum_{c \in \Ic} N_{ab}^{~c}  \, c ~.
\labl{eq:unpert-op}

Defect lines can form junctions, for example when fusing two defects not along their entire length, but only a along a segment. (A defect junction can alternatively be thought of as an insertion of a `defect-joining field' of left/right conformal dimension 0.) The space of possible couplings joining two incoming defects $a$ and $b$ to an out-going defect $c$ is $N_{ab}^{~c}$-dimensional \cite{defect}. The same holds when the roles of in- and out-going defects are reversed. In the nonzero coupling spaces (i.e.\ if $N_{ab}^{~c} = 1$) we choose, once and for all, basis elements such that
\be
  \raisebox{-25pt}{
  \begin{picture}(102,58)
   \put(0,0){\scalebox{.75}{\includegraphics{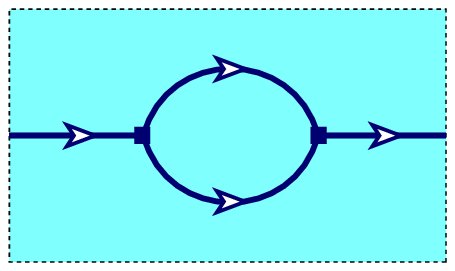}}}
   \put(0,0){
     \setlength{\unitlength}{.75pt}\put(-34,-15){
     \put( 41, 56) {\scriptsize$ c $}
     \put(149, 56) {\scriptsize$ c $}
     \put(112, 71) {\scriptsize$ a $}
     \put(117, 30) {\scriptsize$ b $}
     }\setlength{\unitlength}{1pt}}
  \end{picture}}
  ~=~
  \raisebox{-25pt}{
  \begin{picture}(102,58)
   \put(0,0){\scalebox{.75}{\includegraphics{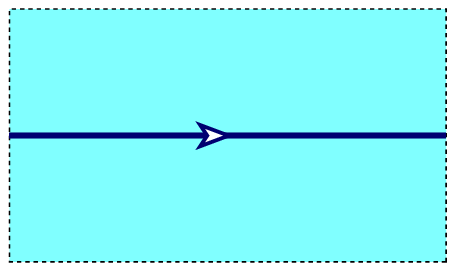}}}
   \put(0,0){
     \setlength{\unitlength}{.75pt}\put(-34,-15){
     \put( 70, 56) {\scriptsize$ c $}
     }\setlength{\unitlength}{1pt}}
  \end{picture}}
  \quad .
\labl{eq:no-bubble}
In words, a `defect bubble' without defect field insertions and which does not enclose any bulk fields can be omitted from the defect line. The identity \erf{eq:no-bubble}, as well as similar identities below, are valid locally on the surface under consideration in the sense that if the left hand side appears as part of a correlator, it can be replaced by the right hand side without affecting the value of the correlator. Next, when fusing two defects along a segment one has the identity \cite{defect}
\be
  \raisebox{-25pt}{
  \begin{picture}(88,58)
   \put(0,0){\scalebox{.75}{\includegraphics{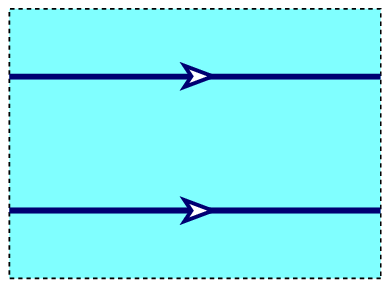}}}
   \put(0,0){
     \setlength{\unitlength}{.75pt}\put(-88,-12){
     \put(120, 75) {\scriptsize$ a $}
     \put(120, 36) {\scriptsize$ b $}
     }\setlength{\unitlength}{1pt}}
  \end{picture}}
  ~=~ \sum_{c \in \Ic} ~
  \raisebox{-25pt}{
  \begin{picture}(88,58)
   \put(0,0){\scalebox{.75}{\includegraphics{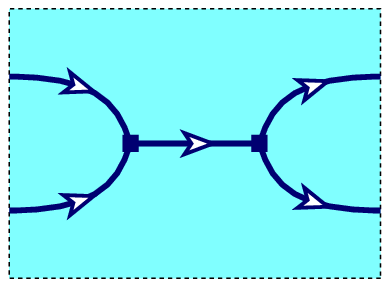}}}
   \put(0,0){
     \setlength{\unitlength}{.75pt}\put(-88,-12){
     \put( 95, 75) {\scriptsize$ a $}
     \put( 95, 37) {\scriptsize$ b $}
     \put(182, 75) {\scriptsize$ a $}
     \put(182, 37) {\scriptsize$ b $}
     \put(143, 57) {\scriptsize$ c $}
     }\setlength{\unitlength}{1pt}}
  \end{picture}}
  \quad .
\ee
Here it is understood that the coupling of the defects $a$ and $b$ to $c$ is zero if $N_{ab}^{~c}=0$.

The space $\Hc^{a\leftarrow b}$ of defect fields that change a defect of type $b$ to a defect of type $a$ decomposes into representations of  $\Vc \otimes \bar \Vc$ as \cite{Petkova:2000ip,Petkova:2001ag,tft1,tft4}
\be
  \Hc^{a \leftarrow b} = \bigoplus_{i,j\in\Ic} \big( R_i \otimes \bar R_j 
  \big)^{\oplus(\sum_{c\in \Ic} N_{ia}^{~c} \, N_{cj}^{~b}) } ~.
\labl{eq:deffield-space}
The space of bulk fields \erf{eq:H-space} is then the space of defect fields living on the invisible defect (labelled by $R_0$), so that $\Hc = \Hc^{0 \leftarrow 0}$. 

Let us choose, once and for all, for each pair $a,b$ a particular defect field $\phi^{a\leftarrow b} \in \Hc^{a \leftarrow b}$. For the construction below to work this choice cannot be arbitrary, but is subject to the following restrictions. First of all, $\phi^{a\leftarrow b}$ has to be chiral. Second, we want all $\phi^{a\leftarrow b}$ to transform in the same representation. So let us fix a preferred representation label $f \in \Ic$ and demand that $\phi^{a\leftarrow b}$ is an element of the sector $R_f \otimes \bar R_0 \subset \Hc^{a \leftarrow b}$. If that sector does not appear in $\Hc^{a \leftarrow b}$ for a particular choice of $a,b$ we set $\phi^{a\leftarrow b}$ to zero. Finally, all $\phi^{a\leftarrow b}$ have to be proportional to the same ground state vector\footnote{
  While for minimal model the space of ground states in an 
  irreducible representation is 
  one-dimensional (and given by multiples of the highest 
  weight vector), for example for a 
  $\widehat{\su}(2)_k$-representation of
  spin $j$ it is $2j{+}1$ dimensional.}
in $R_f \otimes \bar R_0$. That is, we pick a vector $|f\rangle \in R_f$ which is annihilated by all positive modes and demand
\be
  \phi^{a\leftarrow b} \propto \, |f\rangle{\otimes}|\bar 0\rangle
  \quad \in R_f \otimes \bar R_0 \subset \Hc^{a \leftarrow b}
  ~,
\labl{eq:phi-prop}
where $|0\rangle \in R_0$ denotes the vacuum state of $\Vc$. The reason for this last restriction is that the mechanism leading to the functional relation \erf{eq:mm-fus-2s} will require two defect changing fields to sum to zero, which is only possible if they are proportional. The sum of two defect changing fields appears because two defects are fused in \erf{eq:mm-fus-2s}. If one were to look for functional relations involving the fusion of three or more defects, this proportionality is no longer required and condition \erf{eq:phi-prop} should be dropped.

The OPE of defect fields can be computed using the TFT approach as in \cite{tft4}. Rather than reviewing the details, I will quote some results below and sketch the relevant calculations in appendix \ref{app:tft-calc}. (As an aside, an immediate implication of the TFT approach is that in the Cardy case one can use the same OPE coefficients for chiral defect fields as for boundary fields, and the latter are known from \cite{Runkel:1998pm}.) It is convenient to fix the normalisation of the defect fields in terms of constants $\eta^{ab} \in \Cb$ such that
\be
  \phi^{a\leftarrow b}(x) \,  \phi^{b\leftarrow a}(0)
  ~=~ \eta^{ab} \, \eta^{ba} \, \Fs^{(ffa)a}_{b0} \cdot
  x^{-2 h_f} \, \one^{a\leftarrow a} ~+~ \text{(other)} ~~,
\labl{eq:def-norm}
where $h_f$ is the conformal weight of the chosen ground state $|f\rangle \in R_f$ and $\one^{a\leftarrow a}$ is the identity field\footnote{
  In fact, instead of $x^{-2 h_f} \,\one$ the correct expression
  actually is $V^0_{ff}(|f\rangle,x)|f\rangle = C \, x^{-2 h_f} \,\one + \cdots$
  where $V^0_{ff}$ is an intertwiner 
  $R_f \times R_f \rightarrow R_0$ and the 
  constant $C$ depends on the choice of $V^0_{ff}$ and $|f\rangle$.
  For minimal models one takes $|f\rangle$ to be the highest weight vector 
  and normalises the intertwiners such that $C=1$, but e.g.\ for
  $\widehat{\su}(2)_k$ this is not a natural thing to do since
  the space of ground states in $R_f$ need not be one-dimensional.
 } 
on the $a$-defect. The constants $\Fs^{(ijk)l}_{pq}$ and $\Gs^{(ijk)l}_{pq}$ (to appear soon) are entries of the fusing matrix and its inverse, respectively, and describe the transformation behaviour of four-point conformal blocks. They appear in abundance when evaluating expressions obtained in the TFT approach, and are briefly reviewed in appendix \ref{app:fus-def}. Explicit expressions for minimal models are given in appendix \ref{app:mm-chiral}.

When collapsing a defect-bubble in the presence of defect fields, one finds the identities (see appendix \ref{app:tft-calc})
\be
  \raisebox{-25pt}{
  \begin{picture}(102,58)
   \put(0,0){\scalebox{.75}{\includegraphics{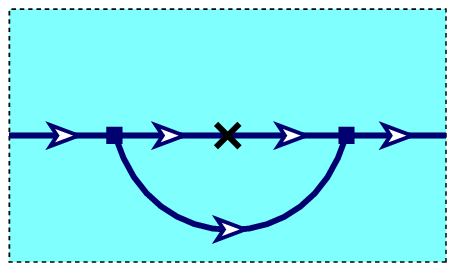}}}
   \put(0,0){
     \setlength{\unitlength}{.75pt}\put(-34,-15){
     \put( 45, 58) {\scriptsize$ d $}
     \put(142, 58) {\scriptsize$ c $}
     \put( 79, 58) {\scriptsize$ b $}
     \put(114, 58) {\scriptsize$ a $}
     \put(117, 23) {\scriptsize$ e $}
     \put( 88, 37) {\scriptsize$ \phi^{a \leftarrow b} $}
     }\setlength{\unitlength}{1pt}}
  \end{picture}}
  ~=~ \frac{\eta^{ab}}{\eta^{cd}} \, \Gs^{(fae)d}_{bc}~
  \raisebox{-25pt}{
  \begin{picture}(102,58)
   \put(0,0){\scalebox{.75}{\includegraphics{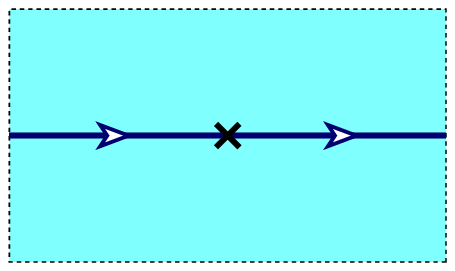}}}
   \put(0,0){
     \setlength{\unitlength}{.75pt}\put(-34,-15){
     \put( 45, 56) {\scriptsize$ d $}
     \put(142, 56) {\scriptsize$ c $}
     \put( 88, 36) {\scriptsize$ \phi^{c \leftarrow d} $}
     }\setlength{\unitlength}{1pt}}
  \end{picture}}
\nonumber  
\ee  
as well as
\be
  \raisebox{-25pt}{
  \begin{picture}(102,58)
   \put(0,0){\scalebox{.75}{\includegraphics{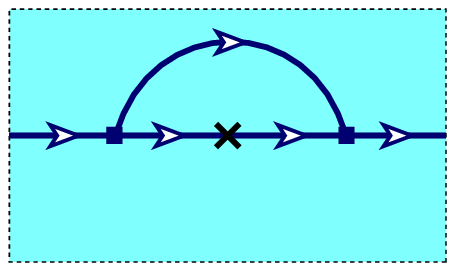}}}
   \put(0,0){
     \setlength{\unitlength}{.75pt}\put(-34,-15){
     \put( 45, 58) {\scriptsize$ d $}
     \put(142, 58) {\scriptsize$ c $}
     \put( 79, 58) {\scriptsize$ b $}
     \put(114, 58) {\scriptsize$ a $}
     \put(117, 76) {\scriptsize$ e $}
     \put( 88, 36) {\scriptsize$ \phi^{a \leftarrow b} $}
     }\setlength{\unitlength}{1pt}}
  \end{picture}}
 ~=~ \frac{\eta^{ab}}{\eta^{cd}} \, \Gs^{(fae)d}_{bc} \,
   \frac{\Rs^{(be)d}}{\Rs^{(ae)c}}~
  \raisebox{-25pt}{
  \begin{picture}(102,58)
   \put(0,0){\scalebox{.75}{\includegraphics{pic04b.eps}}}
   \put(0,0){
     \setlength{\unitlength}{.75pt}\put(-34,-15){
     \put( 45, 56) {\scriptsize$ d $}
     \put(142, 56) {\scriptsize$ c $}
     \put( 88, 36) {\scriptsize$ \phi^{c \leftarrow d} $}
     }\setlength{\unitlength}{1pt}}
  \end{picture}}   
  \quad .
\labl{eq:collapse-field-bubble}
The constants $\Rs^{(ij)k}$ are entries of the braiding matrix and describe how three-point blocks behave under exchange of insertion points. For minimal models these are\footnote{
  This statement depends on the basis of intertwiners
  $R_i \times R_j \rightarrow R_k$ one chooses. So `these are' should
  really be replaced by `there is a basis of intertwiners such 
  that these are'.
  }
just complex phases determined by the conformal weights, cf.\ \erf{eq:mm-Rs-val} in the appendix.

\subsection{Chirally perturbed defects}
\label{sec:chi-pert}

We would now like to perturb a defect of type $a$ by the chiral defect field $\phi^{a\leftarrow a}$. For the fusion procedure in section \ref{sec:pert-fus} below to work without complications, let us assume that the perturbation by $\phi^{a\leftarrow a}$ does not require regularisation, i.e.\ that the leading divergence in the OPE of $\phi^{a\leftarrow a}(x)$ and $\phi^{a\leftarrow a}(y)$ is less singular than $(x-y)^{-1}$. Comparing to \erf{eq:def-norm}, this implies in particular\footnote{
  Since we are explicitly allowing non-unitary theories (otherwise
  the minimal models would not yield examples of the construction
  described here), fields with negative weights are possible and
  the coupling to the identity field does not necessarily give
  the leading singularity.
  }
\be
  h_f < \frac12 ~.
\ee
Consider the cylinder $C(L)$ of circumference $L$ obtained by taking the quotient $C(L) = \Cb / \langle z \mapsto z+L \rangle$, cf.\ figure \ref{fig:fused-def}. For $x_1,\dots,x_n \in \Rb$ and $z \in \Cb$ let $D_a(x_1,\dots,x_n;z)$ be the defect $D_a$ placed on the line $\Rb+z$ in $C(L)$ with defect fields $\phi^{a\leftarrow a}$ inserted at the points $z{+}x_1,\dots,z{+}x_n$. The defect $D_a$ perturbed by $\phi^{a\leftarrow a}$ will be denoted by $D_a(\lambda \phi^{a\leftarrow a};z)$ and is obtained by inserting the exponential 
$\exp\!\big(\lambda \int_0^L \phi^{a\leftarrow a}(x+z) dx \big)$ on the defect line. Explicitly, 
\be
  D_a(\lambda \phi^{a\leftarrow a};z)
  \,=\, \sum_{n=0}^{\infty} \frac{\lambda^n}{n!}
  \int_0^L \!\!\! dx_1 \cdots dx_n \,
  D_a(x_1,\dots,x_n;z) ~~.
\labl{eq:Da-pert-expl}
By changing integration parameters we see that 
$D_a(\lambda \phi^{a\leftarrow a};z) = 
D_a(\lambda \phi^{a\leftarrow a};z+x)$ for all $x \in \Rb$.
Furthermore
$\tfrac{\partial}{\partial\bar z} 
D_a(\lambda \phi^{a\leftarrow a};z) = 0$, 
since $\phi^{a\leftarrow a}$ is a chiral field and so
$\tfrac{\partial}{\partial\bar z}$ annihilates each of the summands on the right hand side of \erf{eq:Da-pert-expl}. Combining these two observations, it follows that
\be
  \frac{\partial}{\partial y}
  D_a(\lambda \phi^{a\leftarrow a};iy) = 0
  \qquad ; ~ y\in\Rb ~.
\ee
So as already announced in the introduction, we can move a defect along the cylinder without affecting the correlator under consideration, as long as the defect line does not cross any field insertions or other defect lines. From here on the perturbed defect will just be denoted by 
$D_a(\lambda \phi^{a\leftarrow a})$ instead of
$D_a(\lambda \phi^{a\leftarrow a};z)$.

Note that $D_a(\lambda \phi^{a\leftarrow a})$ still commutes with the anti-holomorphic modes of the chiral algebra,
\be
  [{\overline W}_m, D_a(\lambda \phi^{a\leftarrow a}) ] = 0 ~.
\ee
Due to the simple decomposition \erf{eq:H-space} of the space of states in the Cardy case, and since we know that $D_a(\lambda \phi^{a\leftarrow a})$ preserves the anti-holomorphic representation of each sector $R_k \otimes \bar R_{\bar k}$, it has no choice but to also preserve the holomorphic representation. Thus it maps each sector $R_k \otimes \bar R_{\bar k}$ to itself.

Finally, since $D_a(\lambda \phi^{a\leftarrow a})$ commutes with the anti-holomorphic component $\bar T$ of the stress-tensor one can easily compute the reflection and transmission coefficients of the defect as defined\footnote{
  In \cite{Quella:2006de} these coefficients were only 
  defined for conformal
  defects. For non-conformal defects (with still critical bulk) one
  should place the defining correlator given in 
  \cite[eqn.\,(2.10)]{Quella:2006de} on a
  cylinder and take the limit where the insertion points are far 
  from the defect. For chirally perturbed defects that does
  not make a difference, since the correlators are identically zero.
  A related quantity defined directly for off-critical (and also
  critical) bulk and defect is the `entropic admittance' introduced 
  in \cite{Friedan:2005ca}.
  }
in \cite{Quella:2006de}, resulting in the reflection being $0$ and the transmission being $1$, independent of the value of $\lambda$.

\subsection{Perturbations by defect changing fields}
\label{sec:defch-pert}

On a superposition $a+b$ of defects, apart from perturbing by the defect fields $\phi^{a\leftarrow a}$ and $\phi^{b\leftarrow b}$ one can also perturb by the defect changing fields $\phi^{a\leftarrow b}$ and $\phi^{b\leftarrow a}$. The corresponding defect operator is
\be
  D_{a+b}\big( 
  \lambda_{aa} \phi^{a\leftarrow a}
  + \lambda_{bb} \phi^{b\leftarrow b}
  + \lambda_{ab} \phi^{a\leftarrow b}
  + \lambda_{ba} \phi^{b\leftarrow a} \big) ~.
\ee
When expanding out the exponential as in \erf{eq:Da-pert-expl}, only terms with the same number of $\phi^{a\leftarrow b}$ insertions as $\phi^{b\leftarrow a}$ insertions can contribute. This is so since $\phi^{a\leftarrow b}(x) \phi^{a\leftarrow b}(y) = 0$ and $\phi^{a\leftarrow b}(x) \phi^{a\leftarrow a}(y) = 0$, and hence every $\phi^{a\leftarrow b}$ insertion must at some point (possibly after a number of defect preserving insertions) be paired off with a $\phi^{b\leftarrow a}$ insertion. In particular, if only $\phi^{a\leftarrow b}$ is involved in the perturbation, but not $\phi^{b\leftarrow a}$, {\em no terms} involving the defect changing field can contribute to the expansion of the exponential in the perturbed operator. Thus we have the identity
\be
  D_{a+b}\big( 
  \lambda_{aa} \phi^{a\leftarrow a}
  + \lambda_{bb} \phi^{b\leftarrow b}
  + \lambda_{ab} \phi^{a\leftarrow b} \big)
  ~=~
  D_{a+b}\big( 
  \lambda_{aa} \phi^{a\leftarrow a}
  + \lambda_{bb} \phi^{b\leftarrow b} \big) ~.
\labl{eq:def-split1}
Since the right hand side contains no contribution mixing the two defects, the perturbed operator is just the sum of the two individual perturbations,
\be
  D_{a+b}\big( 
  \lambda_{aa} \phi^{a\leftarrow a}
  + \lambda_{bb} \phi^{b\leftarrow b} \big) 
  ~=~
  D_a\big( 
  \lambda_{aa} \phi^{a\leftarrow a}\big) 
  + D_b\big(\lambda_{bb} \phi^{b\leftarrow b} \big) ~.
\labl{eq:def-split2}

That a perturbation of a superposition of elementary defects by a defect changing field $\phi^{a\leftarrow b}$ without its partner $\phi^{b\leftarrow a}$ does not affect the defect operator was already noted in \cite{Fuchs:2007tx}. It is also pointed out there that the defect condition itself {\em does} change under the perturbation (in the example considered there, the twisted Hamiltonian becomes non-diagonalisable). It is only the defect {\em operator} that is unaffected.

\subsection{Fusion of chirally perturbed defects}
\label{sec:pert-fus}

As we have seen in section \ref{sec:chi-pert}, correlators on $C(L)$ are independent of the precise location of a chirally perturbed defect. One can thus insert a defect circle of type $a$, perturbed by $\lambda \phi^{a\leftarrow a}$, and another of type $b$, perturbed by $\mu \phi^{b\leftarrow b}$, at a finite distance from each other and take the limit of vanishing distance without encountering a singularity (see figure \ref{fig:fused-def}). In other words, the composition of the chirally perturbed defect operators 
$D_a(\lambda \phi^{a\leftarrow a})$ and 
$D_b(\mu \phi^{b\leftarrow b})$ is well-defined. 
Suppose that the unperturbed defects $a$ and $b$ fuse to $a \star b = c_1+\cdots+c_n$.
To compute the result of the composition, we expand out the exponentials generating the two perturbations and use identities of the form
\be
  \raisebox{-42pt}{
  \begin{picture}(170,95)
   \put(0,0){\scalebox{.75}{\includegraphics{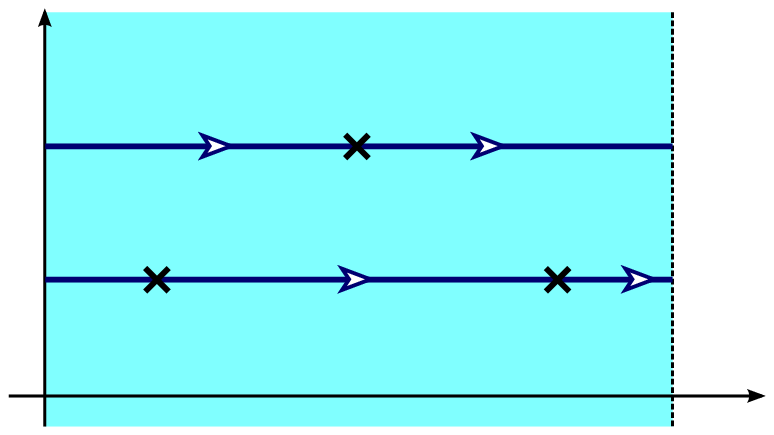}}}
   \put(0,0){
     \setlength{\unitlength}{.75pt}\put(-140,-6){
     \put(142,  4) {\scriptsize$ 0 $}
     \put(335,  4) {\scriptsize$ L $}
     \put(176, 33) {\scriptsize$ \phi^{b\leftarrow b} $}
     \put(290, 33) {\scriptsize$ \phi^{b\leftarrow b} $}
     \put(235, 96) {\scriptsize$ \phi^{a\leftarrow a} $}
     \put(250, 38) {\scriptsize$ b $}
     \put(155, 38) {\scriptsize$ b $}
     \put(296, 92) {\scriptsize$ a $}
     }\setlength{\unitlength}{1pt}}
  \end{picture}}
  =~~ \sum_{i,j,k=1}^n
  \raisebox{-42pt}{
  \begin{picture}(170,95)
   \put(0,0){\scalebox{.75}{\includegraphics{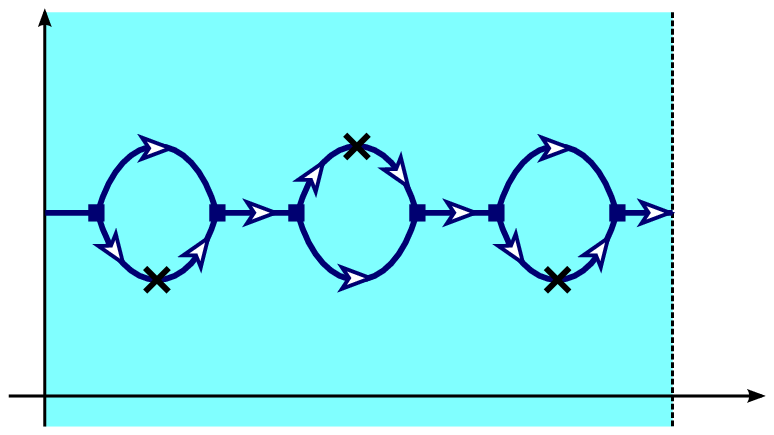}}}
   \put(0,0){
     \setlength{\unitlength}{.75pt}\put(-140,-6){
     \put(142,  4) {\scriptsize$ 0 $}
     \put(335,  4) {\scriptsize$ L $}
     \put(153, 74) {\scriptsize$ c_i $}
     \put(209, 74) {\scriptsize$ c_j $}
     \put(267, 74) {\scriptsize$ c_k $}
     \put(323, 74) {\scriptsize$ c_i $}
     \put(176, 33) {\scriptsize$ \phi^{b\leftarrow b} $}
     \put(290, 33) {\scriptsize$ \phi^{b\leftarrow b} $}
     \put(235, 96) {\scriptsize$ \phi^{a\leftarrow a} $}
     \put(163, 50) {\scriptsize$ b $}
     \put(199, 50) {\scriptsize$ b $}
     \put(279, 50) {\scriptsize$ b $}
     \put(315, 50) {\scriptsize$ b $}
     \put(240, 38) {\scriptsize$ b $}
     \put(182, 92) {\scriptsize$ a $}
     \put(257, 82) {\scriptsize$ a $}
     \put(221, 82) {\scriptsize$ a $}
     \put(296, 92) {\scriptsize$ a $}
     }\setlength{\unitlength}{1pt}}
  \end{picture}}
\labl{eq:pert-fus-id1}
on each term. One can then apply \erf{eq:collapse-field-bubble} to collapse each of the defect bubbles to obtain the appropriate defect (changing) field. For example, collapsing the three bubbles in the example \erf{eq:pert-fus-id1} results in insertions of, in the same order as in \erf{eq:pert-fus-id1},
\be
  \frac{\eta^{bb}}{\eta^{c_jc_i}}
  \Gs^{(fba)c_i}_{bc_j} \frac{\Rs^{(ba)c_i}}{\Rs^{(ba)c_j}} \cdot
  \phi^{c_j\leftarrow c_i}
  \quad , \quad
  \frac{\eta^{aa}}{\eta^{c_kc_j}}
  \Gs^{(fab)c_j}_{ac_k} \cdot
  \phi^{c_k\leftarrow c_j}
  \quad , \quad
  \frac{\eta^{bb}}{\eta^{c_ic_k}}
  \Gs^{(fba)c_k}_{bc_i} \frac{\Rs^{(ba)c_k}}{\Rs^{(ba)c_i}}  \cdot
  \phi^{c_i\leftarrow c_k}
  ~.
\ee
Altogether we find that
\be
D_a\big(\lambda \phi^{a\leftarrow a}\big) \,
D_b\big(\mu \phi^{b\leftarrow b}\big) ~=~
D_{c_1+\cdots+c_n}\big( {\textstyle
\sum_{i,j=1}^n} \, \xi_{ij} \, \phi^{c_i\leftarrow c_j} \big) ~,
\labl{eq:pert-defect-prod}
where
\be
  \xi_{ij} ~=~
  \lambda \cdot 
  \frac{\eta^{aa}}{\eta^{c_ic_j}}
  \Gs^{(fab)c_j}_{ac_i}  
  ~+~
  \mu \cdot
  \frac{\eta^{bb}}{\eta^{c_ic_j}}
  \Gs^{(fba)c_j}_{bc_i}
  \frac{\Rs^{(ba)c_j}}{\Rs^{(ba)c_i}}
\ee
In this computation we have implicitly used the fact that the perturbation does not require regularisation, so that contributions in \erf{eq:pert-fus-id1} with coinciding insertion points have zero weight in the integral. Also, as usual it is understood that the $\Gs$-entries which are not allowed by the fusion rules are set to zero. (This boils down to multiplying $\xi_{ij}$ by $N_{fc_i}^{~c_j}$.)

\subsection{Commuting defect operators}

Recall from \erf{eq:unpert-op} that the unperturbed defect operators act as a multiple of the identity on each sector $R_k \otimes \bar R_{\bar k}$ of the space of states. Consequently they all commute amongst each other,
\be
  [D_a,D_b] = 0 \qquad
  \text{for all} ~
  a,b \in \Ic ~~.
\labl{eq:un-un-comm}
As an aside it is worth pointing out that all defects preserving $\Vc \otimes \bar\Vc$ commute if and only if $Z_{ij} \in \{0,1\}$ for all entries of the modular matrix $Z$ specifying the decomposition of the space of bulk states \cite{BEK,Ost,Fuchs:2007vk}. 

Since chirally perturbed defect operators map each sector $R_k \otimes \bar R_{\bar k}$ to itself, \erf{eq:un-un-comm} continues to hold if only one of the two defects is perturbed,
\be
  [D_a,D_b(\lambda \phi^{b\leftarrow b})] = 0 \qquad
  \text{for all} ~
  a,b \in \Ic ~,~ \lambda \in \Cb ~.
\labl{eq:un-pert-comm}

Two perturbed defects will in general not commute since both defect operators no longer act as a multiple of the identity on each sector $R_k \otimes \bar R_{\bar k}$. However, as announced in \erf{eq:mm-commute}, we will see that in special cases there can be exceptions.

\subsection{Defect fusion with one channel}

Suppose that two defects $a$ and $b$ fuse to a single elementary defect $a \star b = c$ rather than to a superposition. Suppose further that $b$ allows for a chiral defect field in representation $R_f$. Then from \erf{eq:pert-defect-prod} we can read off
\be
D_a \,
D_b\big(\lambda \phi^{b\leftarrow b}\big) =
D_{c}\big(  \xi  \phi^{c\leftarrow c} \big) 
\quad , ~~
  \xi =
  \lambda \cdot
  \frac{\eta^{bb}}{\eta^{cc}}
  \Gs^{(fba)c}_{bc}  ~.
\labl{eq:fus-1-ch}
This relation will be needed when discussing the examples further below.

\subsection{Defect fusion with two channels}

Suppose that two defects $a$ and $b$ fuse to two different elementary defects, $a \star b = c + d$, and that both defects $a$ and $b$ allow for a chiral defect field in representation $R_f$.
Then from \erf{eq:pert-defect-prod}
\be
D_a\big(\lambda \phi^{a\leftarrow a}\big) \,
D_b\big(\mu \phi^{b\leftarrow b}\big) ~=~
D_{c+d}\big( 
  \xi_{cc} \phi^{c\leftarrow c} 
  + \xi_{dd} \phi^{d\leftarrow d} 
  + \xi_{cd} \phi^{c\leftarrow d} 
  + \xi_{dc} \phi^{d\leftarrow c} \big)
\labl{eq:two-channel}
where the couplings to the defect changing fields are given
by
\be
  \xi_{cd} =
  \lambda \cdot 
  \frac{\eta^{aa}}{\eta^{cd}}
  \Gs^{(fab)d}_{ac}
  +
  \mu \cdot
  \frac{\eta^{bb}}{\eta^{cd}} \Gs^{(fba)d}_{bc} 
  \frac{\Rs^{(ba)d}}{\Rs^{(ba)c}} 
  ~~ , ~~
  \xi_{dc} =
  \lambda \cdot 
  \frac{\eta^{aa}}{\eta^{dc}}
  \Gs^{(fab)c}_{ad} 
  + \mu \cdot
  \frac{\eta^{bb}}{\eta^{dc}} \Gs^{(fba)c}_{bd} 
  \frac{\Rs^{(ba)c}}{\Rs^{(ba)d}} 
  ~.
\labl{eq:two-channel-xi}
Provided that $\Gs^{(fba)d}_{bc}$ or $\Gs^{(fba)c}_{bd}$ are non-zero,
we can now set $\xi_{cd}=0$ or $\xi_{dc}=0$ by choosing $\mu$ appropriately. Say for $\mu = \mu^+(\lambda)$ we have $\xi_{cd}=0$ and for $\mu = \mu^-(\lambda)$ we have $\xi_{dc}=0$. Substituting this into $\xi_{cc}$ and $\xi_{dd}$ determines these constants solely in terms of $\lambda$. Let us denote the resulting functions as
$\xi_{cc}^\pm(\lambda)$ and $\xi_{dd}^\pm(\lambda)$. 
Combining \erf{eq:def-split1} and \erf{eq:def-split2} we
finally obtain the identity, for $\eps = \pm$,
\be
D_a\big(\lambda \phi^{a\leftarrow a}\big) \,
D_b\big(\mu^\eps(\lambda) \phi^{b\leftarrow b}\big) ~=~
D_c\big( 
  \xi_{cc}^\eps(\lambda) \phi^{c\leftarrow c} \big) 
+
D_d\big( 
  \xi_{dd}^\eps(\lambda) \phi^{d\leftarrow d} \big) ~.
\labl{eq:proto-rel}
This is the prototypical functional relation for the 
chirally perturbed defect operators which we will use
in the investigations below. If e.g.\ the defect $c$
does not allow for a chiral defect field in representation
$R_f$, \erf{eq:proto-rel} still remains valid, but with
$D_c(\xi_{cc}^\pm(\lambda) \phi^{c\leftarrow c})$ replaced by
the unperturbed defect $D_c$. The same holds for the defect $d$.

To reiterate a remark from the introduction, finding
this relation only relied on the fact that the 
unperturbed defects $a$ and $b$ fuse to a superposition
of {\em two} elementary defects. If $a \star b$ decomposes into
more than two summands, fixing $\mu$ in terms of $\lambda$
will generally not suffice to remove enough defect changing
fields to allow us to split the perturbed defect operator
$D_{c_1+c_2+c_3+\cdots}(\cdots)$ into a sum of individual
defect operators.

\subsection{Defects with su(2)-type fusion}
\label{sec:su2-fus}

Suppose there exists a subset $\{ (m) \,|\, m{=}0,1,\dots,k \}$ of 
elementary defects with $\widehat{\su}(2)_k$-fusion rules,
\be
  (m) \star (n) = \sum_{s=|m-n|,2}^{\min(m+n,2k-m-n)} (s)  ~~,
\ee
where the `$,2$' means that the sum is taken in steps of two.
Suppose further that the defects $(m)$ with $m=1,2,\dots,k{-}1$ support a chiral defect field in representation $R_f$. Since the fusion of defect lines (in the Cardy case) agrees with that of the irreducible representations labelling the defect, this implies $f=(2)$. Furthermore, since
\be
  D_{(1)} \, D_{(s)} = D_{(s-1)} + D_{(s+1)}
  \qquad \text{for}~s=1,2,\dots,k{-}1~~,
\ee
we can apply \erf{eq:proto-rel}. For example, for $s=1$ one
finds
\be
  D_{(1)}\big(\lambda \phi^{1\leftarrow1}\big) \,
  D_{(1)}\big(\mu \phi^{1\leftarrow1}\big)
  = D_{(0)+(2)}\big( 
  \xi_{02} \phi^{0\leftarrow2} +
  \xi_{20} \phi^{2\leftarrow0} +
  \xi_{22} \phi^{2\leftarrow2}\big) ~,
\labl{eq:su2-11-fus}
($\phi^{0\leftarrow0}$ is zero) with
\be
   \xi_{02} = \frac{\eta^{11}}{\eta^{02}} \Gs^{(211)2}_{10}
   \big( \lambda - \omega^{-1} \mu  \big) ~,~~
   \xi_{20} = \frac{\eta^{11}}{\eta^{20}} \Gs^{(211)0}_{12}
   \big( \lambda - \omega \mu \big) ~,~~
   \xi_{22} = \frac{\eta^{11}}{\eta^{22}} \Gs^{(211)2}_{12}
   \big( \lambda + \mu \big) ~,
\ee  
where $\omega = -\Rs^{(11)0} / \Rs^{(11)2} \neq 0$. (For $N_{ij}^{~k}=1$, $\Rs^{(ij)k}$ describes a basis transformation of a one-dimensional space and hence is never zero.) From \erf{eq:su2-11-fus} we can learn two things. First, setting
$\mu = \omega^{\pm 1} \lambda$ results in the functional
relation
\be
  D_{(1)}\big(\lambda \phi^{1\leftarrow1}\big) \,
  D_{(1)}\big(\omega^{\pm 1} \lambda \phi^{1\leftarrow1}\big)
  = \id + D_{(2)}\Big( 
  \frac{\eta^{11}}{\eta^{22}} \Gs^{(211)2}_{12}
   \big(1 + \omega^{\pm1}\big) \lambda
  \phi^{2\leftarrow2}\Big) ~,
\ee
Second, if one expands out the exponential generating the perturbation in \erf{eq:su2-11-fus}, as remarked already in section \ref{sec:defch-pert}, for each insertion $\phi^{0\leftarrow2}$ there has to be a corresponding insertion of $\phi^{2\leftarrow0}$. Thus for all non-vanishing terms in the expansion the coefficients $\xi_{02}$ and $\xi_{20}$ only appear in the combination $\xi_{02}\xi_{20} = \text{(const)}\big(\lambda^2 - (\omega{+}\omega^{-1})\lambda \mu + \mu^2\big)$, which is symmetric under exchange of $\lambda$ and $\mu$. Since also $\xi_{22}$ is invariant under $\lambda \leftrightarrow \mu$ we see that
\be
  D_{(1)}\big(\lambda \phi^{1\leftarrow1}\big) \,
  D_{(1)}\big(\mu \phi^{1\leftarrow1}\big)
  =
  D_{(1)}\big(\mu \phi^{1\leftarrow1}\big) \,
  D_{(1)}\big(\lambda \phi^{1\leftarrow1}\big)
  \qquad \text{for all}~~\lambda,\mu \in \Cb~.
\labl{eq:su2-11-comm}
Under the assumption (which will be checked in the examples below) that there are no `accidental zeros' $\xi(\lambda)=0$ in applying the relation \erf{eq:proto-rel}, starting from the above result one can prove by induction that in fact all of the perturbed defect operators mutually commute for arbitrary values of the coupling constants,
\be
  \big[
  D_{(m)}\big(\lambda \phi^{1\leftarrow1}\big) ,
  D_{(n)}\big(\mu \phi^{1\leftarrow1}\big) \big] = 0
  \qquad \text{for}~
  m,n=1,2,\dots,k{-}1~\text{and}~
  \lambda,\mu \in \Cb~.
\labl{eq:su2-all-comm}
For $m=0$ or $m=k$ this also holds due to \erf{eq:un-pert-comm}. The induction argument is as follows. From \erf{eq:un-pert-comm} and \erf{eq:su2-11-comm} we know that \erf{eq:su2-all-comm} holds for $m,n \le 1$. Suppose now that we have already proved \erf{eq:su2-all-comm} for $m,n \le M$. 
By \erf{eq:proto-rel} there are constants $a,b,c \in \Cb$ such that
(omitting the $\phi^{1\leftarrow1}$ for brevity)
\be
  D_{(1)}(\lambda) 
  D_{(m)}(a \lambda) 
  =
  D_{(m-1)}(b \lambda) +
  D_{(m+1)}(c \lambda) ~~.
\ee
Then by induction assumption, for $m,n \le M$,
\bea
  \big[ D_{(1)}(\lambda) 
  D_{(m)}(a \lambda),
  D_{(n)}(\mu) \big]
 \enl
  \qquad = D_{(1)}(\lambda)
  \big[   D_{(m)}(a \lambda),
  D_{(n)}(\mu) \big]
  + 
  \big[ D_{(1)}(\lambda) ,
  D_{(m)}(a \lambda) \big]
  D_{(n)}(\mu) =0 ~.
\eear\ee
On the other hand,
\bea
  \big[ D_{(1)}(\lambda) 
  D_{(m)}(a \lambda),
  D_{(n)}(\mu) \big]
 \enl
  \qquad = \big[   D_{(m-1)}(b \lambda) +
  D_{(m+1)}(c \lambda) ,
  D_{(n)}(\mu) \big]
  = 0 + \big[ D_{(m+1)}(c \lambda) ,
  D_{(n)}(\mu) \big] ~~.
\eear\ee
Provided that $c \neq 0$ (this is the assumption that there are no `accidental zeros') we thus get $[ D_{(m)}(\lambda) ,
  D_{(n)}(\mu) ] = 0$ for $m \le M{+}1$, $n \le M$ and all $\lambda,\mu\in\Cb$. Running the above argument again with $n=M{+}1$ then shows that \erf{eq:su2-all-comm} also holds for $m,n \le M{+}1$.

\sect{Examples}\label{sec:ex}

\subsection{Virasoro minimal models}\label{sec:ex-mm}

Consider the A-series Virasoro minimal model $M(p,p')$. It has central charge
\be
  c = 13-6\big(t+t^{-1}\big)
  \qquad ; ~~ t = p/p' ~.
\labl{eq:mm-cent}
The irreducible representations $R_{r,s}$ of the Virasoro (vertex) algebra at that central charge are labelled by entries $(r,s)$ of the Kac-table with $1 \,{\le}\, r \,{<}\, p$ and $1 \,{\le}\, s \,{<}\, p'$. The conformal weight of the highest weight state in the representation $R_{r,s}$ is
\be
  h_{r,s} = \big( (d_{r,s})^2 - (1{-}t)^2 \big) / (4t)
  \qquad ; ~~ d_{r,s} = r - st ~.
\labl{eq:hrs-drs}
The labels $(r,s)$ and $(p{-}r,p'{-}s)$ denote the same representation. We choose the distinguished representation to be
\be
  f = (1,3) ~.
\labl{eq:mm-dist}
We then have $h_f = {-}1{+}2t$ and so the condition $h_f< \tfrac12$ amounts to $t < \tfrac34$. 

As described in section \ref{sec:fun-eq}, defects are labelled by irreducible representations, in this case by Kac-labels $(r,s)$ modulo the identification $(r,s) \sim (p{-}r,p'{-}s)$. There are two subsets with $\su(2)$-type fusion rules, namely the defects labelled by $(r,1)$ and by $(1,s)$. Consider the subset $\{(1,s)\}$ first. The identification with the notation in section \ref{sec:su2-fus} is $(m)=(1,m{+}1)$. As discussed there, the only non-trivial chiral perturbation possible for the defect labelled $(1)$ is $f=(2)$, which is precisely the choice \erf{eq:mm-dist}. The same reasoning for the subset $\{(r,1)\}$ would give $f=(3,1)$, but for a given value of $t$ only one of the two is a relevant perturbation (in particular, they can never both obey $h < \tfrac12$), and so we choose $p$ and $p'$ such that $f=(1,3)$ is relevant.

It is convenient to fix the normalisation constant of the defect field $\phi^{(r,s) \leftarrow (r,s)}$ to be
\be
  \eta^{(r,s)(r,s)} = 
  \frac{ \Ga{t}\Ga{2{-}3t} }{
         \Ga{1{-}t{+}d_{r,s}}\Ga{1{-}t{-}d_{r,s}} } ~.
\labl{eq:mm-eta_rs_rs}
Substituting this into \erf{eq:def-norm} and using \erf{eq:F(33x)}
results in, with $\phi \equiv \phi^{(r,s) \leftarrow (r,s)}$
and $d \equiv d_{r,s}$,
\be
  \phi(x)\phi(0)
  =
  \frac{
  \sin\!\big(\pi(t{-}d)\big) \sin\!\big(\pi(t{+}d)\big)
  }{
  \sin(\pi t) \sin(3\pi t)
  } \,
  \frac{ \Ga{2{-}3t} \, \Ga{2t} }{\Ga{2{-}2t}\,\Ga{t}}
  \cdot x^{-2 h_f} \, \one^{a\leftarrow a} ~+~ \text{(other)} ~.
\ee
Next we apply \erf{eq:two-channel} for $a=(1,2)$ and $b=(1,s)$. Substituting \erf{eq:mm-eta_rs_rs}, \erf{eq:mm-Rs-val}, \erf{eq:mm-F-2} and \erf{eq:mm-GF} into \erf{eq:two-channel-xi}, after a while one finds, for $\nu = \pm$ and $d \equiv d_{1,s}$,
\bea
  \xi_{(1,s+\nu),(1,s+\nu)} =
  \lambda \cdot
  \frac{\sin( \pi t )}{\sin(\pi \nu d)}
  + 
  \mu \cdot 
  \frac{\sin( \pi(t{+}\nu d))}{\sin(\pi \nu d)}
  ~~,
\enl
  \xi_{(1,s+\nu),(1,s-\nu)} = 
  \frac{1}{\eta^{(1,s+\nu),(1,s-\nu)}} \,
  \frac{\Ga{2{-}3t}\,\Ga{\nu d}}{\Ga{1{-}t{+}\nu d}\,
  \Ga{1{-}2t}} \,
  \Big( \lambda - \mu \cdot e^{i \pi \nu (1-d)} \Big)~~.
\eear\ee
So if we set $\mu = \lambda \, e^{i \pi \eps (1-d)}$ then for $\eps\,{=}\,1$ we have $\xi_{(1,s-1),(1,s+1)}=0$ while for $\eps\,{=}\,{-}1$ we get $\xi_{(1,s+1),(1,s-1)}=0$. In either case the defect changing fields do no longer contribute the the perturbed defect operator. Furthermore, for this value of $\mu$ one finds $\xi_{(1,s+\nu),(1,s+\nu)} = \lambda\,e^{i \pi \eps(1-d-\nu t)}$, which results in the functional relation
\be
  D_{(1,2)}\big(\lambda \,\phi\big)
  D_{(1,s)}\big(e^{i \pi \eps (1-d)} \lambda \,\phi\big)
  =
  D_{(1,s-1)}\big(e^{i \pi \eps (1-d+t)}\lambda \,\phi\big)
  +
  D_{(1,s+1)}\big(e^{i \pi \eps (1-d-t)} \lambda \,\phi\big) ~.
\labl{eq:mm-fus2-der}
Substituting further $d=1{-}st$ and $q = e^{i \pi t}$ leads to the relation quoted in \erf{eq:mm-fus-2s}. Since there are no `accidental zeros' in the coupling constants on the right hand side of \erf{eq:mm-fus2-der}, the recursive argument leading to \erf{eq:su2-all-comm} applies and we obtain
\be
  \big[ D_{(1,s)}(\lambda\phi), D_{(1,s')}(\mu\phi) \big] = 0
  \qquad \text{for~all}~~
  s,s'=1,\dots,p'{-}1~,~~\lambda,\mu\in\Cb~.
\labl{eq:mm-1s-com}
As before, in this relation, as well as in \erf{eq:mm-fus2-der}, if the defect does not support the chiral $h_{1,3}$-defect field, $D(\lambda\phi)$ stands for the unperturbed defect operator $D$. The results \erf{eq:mm-fus2-der} and \erf{eq:mm-1s-com} can be extended to all defects $D_{(r,s)}$ as follows. Due to the fusion rule $(r,1) \star (1,s) = (r,s)$ we can apply \erf{eq:fus-1-ch}, and substituting \erf{eq:mm-eta_rs_rs} and \erf{eq:F(rs3)} yields
\be
  D_{(r,1)} \, D_{(1,s)}(\lambda \phi) = 
  D_{(r,s)}( (-1)^{r-1} \lambda \phi ) ~.
\ee
This determines the perturbed operators $D_{(r,s)}(\lambda \phi)$ in terms of the unperturbed operators $D_{(r,1)}$ and the perturbed operators in the subset $\{(1,s)\}$. Note also that by \erf{eq:un-pert-comm} we have $[D_{(r,1)},D_{(1,s)}(\lambda \phi)]=0$. This implies that \erf{eq:mm-1s-com} also holds for general defects, establishing \erf{eq:mm-commute}.

To obtain the functional relation \erf{eq:mm-Tsys} now that we have established \erf{eq:mm-fus-2s} works along the same lines as the corresponding calculation for fused row transfer matrices \cite{KlPe}. First note that setting $s\,{=}\,2$, $\eps\,{=}\,{-}1$ and replacing $\lambda \rightarrow q\lambda$ in \erf{eq:mm-fus-2s} one obtains \erf{eq:mm-Tsys} for $s\,{=}\,2$. If \erf{eq:mm-Tsys} holds for $s \le m$ then on the one hand, 
\bea
  D_{(1,m)}(q^{ m} \lambda \phi)
  D_{(1,2)}(\lambda \phi)
  D_{(1,m+1)}(q^{m+1} \lambda \phi)
\enl
  \quad = \Big(
  D_{(1,m-1)}(q^{m+1} \lambda \phi)
  +
  D_{(1,m+1)}(q^{m-1} \lambda \phi)
  \Big)
  D_{(1,m+1)}(q^{m+1} \lambda \phi) ~~,
\eear\ee
and on the other hand, fusing the second and third defect,
\bea
  D_{(1,m)}(q^{ m} \lambda \phi)
  D_{(1,2)}(\lambda \phi)
  D_{(1,m+1)}(q^{m+1} \lambda \phi)
\enl
  \quad = 
  D_{(1,m)}(q^{ m} \lambda \phi)  
  \Big(
  D_{(1,m)}(q^{m+2} \lambda \phi)
  +
  D_{(1,m+2)}(q^{ m} \lambda \phi)
  \Big) ~~.
\eear\ee
Subtracting the two expressions and using that \erf{eq:mm-Tsys} holds for $s\,{=}\,m$ shows that it also holds for $s\,{=}\,m{+}1$.

The final property of $D_{(1,s)}(\lambda \phi)$ to be derived is the behaviour under reflection $s \rightarrow p'{-}s$. This can be deduced by fusing with the defect $D_{(1,p'-1)}$. Substituting \erf{eq:mm-eta_rs_rs} and \erf{eq:F(ps3)} into \erf{eq:fus-1-ch} gives
\be
  D_{(1,p'-1)} \, D_{(1,s)}(\lambda \phi)
  = D_{(1,p'-s)}( (-1)^p \lambda \phi ) ~.
\ee
The action of $D_{(1,p'-1)}$ on the sector $R_{r,s} \otimes \bar R_{r,s}$ of the space of bulk states is given by \erf{eq:unpert-op} in terms of the modular $S$-matrix, which for minimal models can be found e.g.\ in \cite[ch.\,10]{DiFr},
\be
    D_{(1,p'-1)} \big|_{R_{r,s} \otimes \bar R_{r,s}} = 
    (-1)^{rp'+sp+1} \,
    \id_{R_{r,s} \otimes \bar R_{r,s}} ~~.
\ee
So altogether,
\be
  D_{(1,p'-s)}(\lambda \phi) \big|_{R_{r,s} \otimes \bar R_{r,s}} 
  = (-1)^{rp'+sp+1} \,
  D_{(1,s)}((-1)^{p} \lambda \phi) 
  \big|_{R_{r,s} \otimes \bar R_{r,s}} ~~.    
\ee
For the minimal model $M(2,p')$ one has $r\,{=}\,1$ and $p'$ odd, and this reduces to the reflection property 
$D_{(1,p'-s)}(\lambda \phi) = D_{(1,s)}(\lambda \phi)$
already observed for the {\bf T}-operators in \cite{Bazhanov:1994ft}.

\subsection{Liouville theory}

The central charge of Liouville theory is usually parametrised as $c=1+6Q^2$ with $Q = b+b^{-1}$ and the conformal weight of a highest weight vector as $h_\alpha = \alpha(Q{-}\alpha)$. For example, the Verma modules of conformal weight $h_{-b/2}$ and $h_{-b}$ contain null-vectors at levels 2 and 3, respectively. 

While conformal boundary conditions have been analysed in great detail \cite{Fateev:2000ik,Teschner:2000md,Zamolodchikov:2001ah,Ponsot:2001ng}, topological defects in Liouville theory have so far not been studied. However, given the general pattern that in the Cardy case topological defect lines are labelled in the same way as boundary conditions, it seems a reasonable guess that this will remain true also in Liouville theory. The considerations below are based on this assumption. 

\medskip

There is a discrete family $D_{(m,n)}$ of defects for $m,n \ge 1$ corresponding to the point-like ZZ-boundary conditions, and a one-parameter family $D_\sigma$ of defects corresponding to FZZT-boundary conditions. Assuming further that as in the Cardy case for rational CFTs, the fusion of defect lines agrees with that of the representations labelling the defects, we have $(1,2) \star (\sigma) = (\sigma{-}\tfrac{b}2) + (\sigma{+}\tfrac{b}2)$, and $(1,2) \star (m,n) = (m,n{-}1) + (m,n{+}1)$ for $n\ge 2$. The defect $(1,2)$ is thus a candidate to give rise to functional equations between perturbed defect operators. 

The spectrum of boundary fields on the boundary condition labelled $(1,2)$ consists of the irreducible degenerate representations with labels $(1,1)$ and $(1,3)$ \cite{Zamolodchikov:2001ah}. The representation $(m,n)$ has conformal weight $h_\alpha$ with $\alpha = (1{-}m)b^{-1}+(1{-}n)b$. Since the spectrum of chiral defect fields should coincide with the spectrum of boundary fields, the defect $(1,2)$ supports a chiral defect field of weight $h_{-b} = {-}2b^2{-}1$, which is always less than $\tfrac12$. This chiral defect field is also allowed for all defects $(m,n)$ with $n\,{\ge}\,2$. As for minimal models, the defects $(1,n)$ form a subset with $\su(2)$-type fusion rules, albeit now only truncated from below by $(1,1)$, but not from above. In fact, the relations \erf{eq:mm-fus-2s} and \erf{eq:mm-Tsys} for the perturbed defect operators will hold in precisely the same form as for minimal models if we set $q = e^{- i \pi b^2}$. To see this, note that if we replace 
\be
  t \rightarrow -b^{2}
  \quad , \quad
  d \rightarrow b(2\alpha{-}Q)
\ee  
in \erf{eq:mm-cent} and \erf{eq:hrs-drs} we obtain precisely the expression for the Liouville central charge and the conformal weight of $h_\alpha$. Since the derivation of the $\Fs$-matrix entries \erf{eq:mm-F-2} just relied on the existence a level 2 null vector, the corresponding Liouville expressions are obtained by the same replacement (as can be checked explicitly by comparing to the expression in e.g.\ \cite[app.\,B]{Ponsot:2001ng}). The same holds for the $\Fs$-matrix entry \erf{eq:F(33x)}, as this was obtained from \erf{eq:mm-F-2} and the pentagon identity. The calculations in section \ref{sec:ex-mm} leading to \erf{eq:mm-fus-2s} and \erf{eq:mm-Tsys} will therefore go through in the same way if applied to the $(1,n)$-defects in Liouville theory.

The same reasoning applies to the defect $(2,1)$, which is related to $(1,2)$ via $b \leftrightarrow b^{-1}$. As opposed to the minimal model case, for Liouville theory both the $(1,2)$- and the $(2,1)$-defect have a chiral defect field of weight less than $\tfrac12$. For the $(2,1)$-defect this is the field of weight $h_{-b^{-1}} = {-}2b^{-2}{-}1$.

\medskip

Liouville theory in the presence of $(m,n)$-defects will not be unitary, in the sense that some spectra contain complex conformal weights. For example, the spectrum of open states on a strip with boundary conditions $\sigma$ and $(m,n)$ will have complex conformal weights unless $m\,{=}\,n\,{=}\,1$ \cite{Zamolodchikov:2001ah}. Since this spectrum is the same as that of a strip with boundary conditions $\sigma$ and $(1,1)$ (which by itself has a discrete and real spectrum) on which a $(m,n)$-defect line has been inserted parallel to the boundary, we see that in this case the presence of the $(m,n)$-defect leads to complex weights in the spectrum.

\sect{Conclusions}\label{sec:conc}

In this paper I have shown that in certain cases chirally perturbed defect operators satisfy functional relations. For a subset where the unperturbed defects have $\su(2)$ fusion rules, these functional relations are well-known from the study of integrable models as T-system. The perturbed defects in this subset mutually commute, and so it is possible to choose the common eigenvectors independent of the coupling constants. The eigenvalues can then be computed with the help of the thermodynamic Bethe ansatz. 

The idea leading to the functional relations is simply to adjust the couplings of the two perturbed defects to be fused in such a way that the defect changing fields cannot contribute to the operator of the fused defect.

It is also worth pointing out that while in this paper only perturbations by holomorphic defects fields were considered, the analysis can of course be repeated for anti-holomorphic defect fields. It is immediate from the TFT representation of defect correlators (see section \ref{app:tft-calc}) that in the Cardy case the operators for defects perturbed by a holomorphic defect field $\phi$ will commute with those of defects perturbed by an anti-holomorphic field $\bar\psi$,
\be
  \big[\,D_a(\lambda \phi)\,,\,D_b(\mu \bar \psi)\,\big] = 0
\ee
for all defect types $a,b$, all defect fields $\phi, \bar\psi$, and all coupling constants $\lambda,\mu \in \Cb$. The fusion of a defect perturbed by $\phi$ and a defect perturbed by $\bar\psi$ results in a defect perturbed by a linear combination of $\phi$ and $\bar\psi$. 

\medskip\noindent
Many possible directions for further study remain. Some of them are:
\begin{list}{-}{\topsep .4em \itemsep 0em \leftmargin 1em}

\item In this paper only perturbations that did not require regularisation were considered. An obvious task is to extend the method to include all relevant and marginal perturbations.

\item The defect operators constructed in the minimal model example are close cousins of the {\bf T}-operators of \cite{Bazhanov:1994ft,Bazhanov:1996dr,Bazhanov:1998dq}. In these papers also {\bf Q}-operators are defined, which together with {\bf T} obey Baxter's TQ-relation. An interpretation of the {\bf Q}-operators in terms of defects remains to be found.

\item Here perturbed operators were studied only in the charge-conjugation modular invariant theory for a given rational chiral algebra. However, the methods in \cite{tft4,defect} provide the tools to do the same computation also for other local RCFTs, such as e.g.\ the D-series and the exceptional modular invariants for minimal models and $\widehat{\su}(2)_k$. Systems of functional relations are known for various CFTs, see e.g.\ \cite{Zamolodchikov:1991et,Bazhanov:2001xm,Dorey:2006an}, and one could try to obtain them also with the methods presented here. The generalisation to super-conformal models (cf.\ \cite{Kulish:2005qb}) is another open point.

\item As mentioned a number of times, the functional relations satisfied by the perturbed defect operators make it possible to determine its matrix elements at finite values of the coupling constant (at least numerically). One important application of this is the investigation of boundary flows. It would be interesting to carry out a systematic investigation for rational CFTs such as WZW models, and it might also be possible to study boundary flows in non-compact theories, in particular Liouville theory (see e.g.\ \cite{Teschner:2003qk,Graham:2006gca} for existing results), in this way.

\end{list}

\bigskip\noindent
{\bf Acknowledgements:} I would like to thank G\'erard Watts for illuminating discussions on T-systems and disc amplitudes, and I am grateful to him, as well as to Patrick Dorey, J\"urgen Fuchs, Christoph Schweigert and J\"org Teschner for helpful comments on a draft of this paper. This research was partially supported by the EPSRC First Grant EP/E005047/1, the PPARC rolling grant PP/C507145/1 and the Marie Curie network `Superstring Theory' (MRTN-CT-2004-512194).

\appendix

\sect{Appendix}

\subsection{Fusing matrices}
\label{app:fus-def}

Recall that we assume that the chiral algebra $\Vc$ has a finite number of irreducible representations, indexed by a set $\Ic$, and that for simplicity we also demand the fusion rule coefficients to obey $N_{ij}^{~k} \in \{0,1\}$ for all $i,j,k \in \Ic$. Let us choose a basis of conformal three-point blocks, that is, multilinear maps\footnote{
  Actually, the intertwiners are linear maps from $R_i$ to formal 
  Laurent series $z^{h_k-h_i-h_j} \Hom(R_j,R_k)[\![ z^{\pm 1} ]\!]$,
  but I will not make this explicit in the notation below. More
  details can be found in \cite{FrenBen}.
  }
\be
  V^{k}_{ij}(\,\cdot\,,z) : R_i \times R_j
  \longrightarrow R_k \quad , ~~(v_i,v_j) \mapsto
  V^{k}_{ij}(v_i,z)\,v_j ~~, 
\labl{eq:intertw}
which intertwine the action of $\Vc$ in a suitable way, see e.g.\ \cite{Moore:1989vd,FrenBen} (or \cite[ch.\,5]{tft4}, which uses the same conventions as here).

Fusing matrices describe a change of basis in the space of conformal four-point blocks. Consider four representations $R_i$, $R_j$, $R_k$ and $R_l^*$ (the dual of $R_l$) of the chiral algebra $\Vc$, placed at $z$, $w$, $0$ and $\infty$, respectively. There are several ways to give a basis for the space of conformal four point blocks $R_l^* \times R_i \times R_j \times R_k \rightarrow \Cb$ in terms of the intertwiners \erf{eq:intertw}. Two of them are
\be
  B_1 = \big\{ 
  ~(\varphi_l, v_i, v_j, v_k) \mapsto
  \varphi_l\big(\,V^l_{ip}(v_i,z) \,V^p_{jk}(v_j,w)\,v_k\, 
  \big) ~ \big|\,
  p \in \Ic \big\}
\ee
and
\be
  B_2 = \big\{ 
  ~(\varphi_l, v_i, v_j, v_k) \mapsto
  \varphi_l\big(\,V^l_{qk}\big(V^q_{ij}(v_i,z{-}w)\,v_j,w\big) \,v_k\, 
  \big) ~ \big|\,
  q \in \Ic \big\} ~.
\ee
The fusing matrices $\Fs^{(ijk)l}$ are obtained by expressing basis vectors of $B_1$ in terms of those of $B_2$,
\be
  V^l_{ip}(v_i,z) \,V^p_{jk}(v_j,w)
  = \sum_{q\in\Ic}
  \Fs^{(ijk)l}_{pq} \cdot
  V^l_{qk}\big(V^q_{ij}(v_i,z{-}w)\,v_j,w\big)  ~.
\labl{eq:F-def-inter}
For minimal models it is enough to evaluate this relation on the highest weight states of the corresponding representations. This then results in the usual calculation comparing the asymptotic behaviour of conformal four-point blocks \cite[ch.\,8]{DiFr}.

The definition \erf{eq:F-def-inter} of $\Fs^{(ijk)l}$ can be expressed graphically as
\be
  \raisebox{-42pt}{
  \begin{picture}(70,85)
   \put(0,8){\scalebox{.75}{\includegraphics{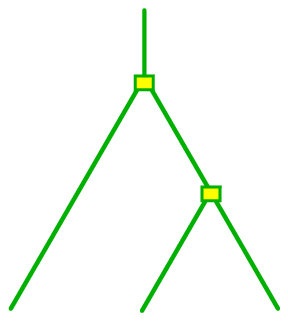}}}
   \put(0,8){
     \setlength{\unitlength}{.90pt}\put(-16,-16){
     \put( 14,  8) {\scriptsize$ i $}
     \put( 47,  8) {\scriptsize$ j $}
     \put( 80,  8) {\scriptsize$ k $}
     \put( 60, 62) {\scriptsize$ p $}
     \put( 48, 95) {\scriptsize$ l $}
     }\setlength{\unitlength}{1pt}}
  \end{picture}}
  = ~ \sum_{q\in\Ic} ~
  \Fs^{(ijk)l}_{pq} 
  \raisebox{-42pt}{
  \begin{picture}(70,85)
   \put(0,8){\scalebox{.75}{\includegraphics{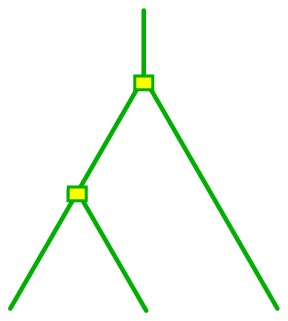}}}
   \put(0,8){
     \setlength{\unitlength}{.90pt}\put(-16,-16){
     \put( 14,  8) {\scriptsize$ i $}
     \put( 47,  8) {\scriptsize$ j $}
     \put( 80,  8) {\scriptsize$ k $}
     \put( 33, 62) {\scriptsize$ q $}
     \put( 48, 95) {\scriptsize$ l $}
     }\setlength{\unitlength}{1pt}}
  \end{picture}}  
  \quad .
\ee
The inverse $\Gs^{(ijk)l}$ of $\Fs^{(ijk)l}$ is expressed accordingly as
\be
  \raisebox{-42pt}{
  \begin{picture}(70,85)
   \put(0,8){\scalebox{.75}{\includegraphics{pic07b.eps}}}
   \put(0,8){
     \setlength{\unitlength}{.90pt}\put(-16,-16){
     \put( 14,  8) {\scriptsize$ i $}
     \put( 47,  8) {\scriptsize$ j $}
     \put( 80,  8) {\scriptsize$ k $}
     \put( 33, 62) {\scriptsize$ q $}
     \put( 48, 95) {\scriptsize$ l $}
     }\setlength{\unitlength}{1pt}}
  \end{picture}}  
  = ~ \sum_{p\in\Ic} ~
  \Gs^{(ijk)l}_{qp} 
  \raisebox{-42pt}{
  \begin{picture}(70,85)
   \put(0,8){\scalebox{.75}{\includegraphics{pic07a.eps}}}
   \put(0,8){
     \setlength{\unitlength}{.90pt}\put(-16,-16){
     \put( 14,  8) {\scriptsize$ i $}
     \put( 47,  8) {\scriptsize$ j $}
     \put( 80,  8) {\scriptsize$ k $}
     \put( 60, 62) {\scriptsize$ p $}
     \put( 48, 95) {\scriptsize$ l $}
     }\setlength{\unitlength}{1pt}}
  \end{picture}}
  \quad .
\ee
The braiding matrix $\Rs^{(ij)k}$ in turn is defined by analytic continuation of three-point blocks and has the graphical representation
\be
  \raisebox{-35pt}{
  \begin{picture}(30,70)
   \put(0,8){\scalebox{.75}{\includegraphics{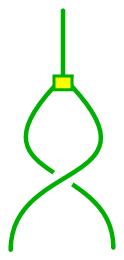}}}
   \put(0,8){
     \setlength{\unitlength}{.90pt}\put(-16,-16){
     \put( 14,  8) {\scriptsize$ i $}
     \put( 40,  8) {\scriptsize$ j $}
     \put( 28, 80) {\scriptsize$ k $}
     }\setlength{\unitlength}{1pt}}
  \end{picture}}
  =~ \Rs^{(ij)k}
  \raisebox{-35pt}{
  \begin{picture}(30,70)
   \put(0,8){\scalebox{.75}{\includegraphics{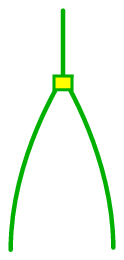}}}
   \put(0,8){
     \setlength{\unitlength}{.90pt}\put(-16,-16){
     \put( 14,  8) {\scriptsize$ i $}
     \put( 40,  8) {\scriptsize$ j $}
     \put( 28, 80) {\scriptsize$ k $}
     }\setlength{\unitlength}{1pt}}
  \end{picture}}
  \quad .
\ee
Some relations between $\Fs$, $\Gs$ and $\Rs$ are collected e.g.\ in \cite[ch.\,2.2]{tft1}.

\subsection{Defect correlators in the TFT approach}
\label{app:tft-calc}

In the TFT approach to rational CFT \cite{Felder:1999cv,Fuchs:2001am}, two-dimensional CFT correlators are expressed in terms of correlators of a three-dimensional topological field theory on a three-manifold with boundary. The relevant manifold is simply given by the surface considered in the CFT correlator times an interval. Field insertions and defect lines are encoded by placing appropriate Wilson lines (``ribbons'') inside this three-manifold. The treatment of defects is described in detail in \cite{tft4,defect}. For example, the ribbon graph corresponding to an insertion of a chiral defect field in representation $R_f$ is\footnote{
  That the defect lines on the surface for the CFT and the ribbons
  inside the three-manifold for the TFT have opposite orientation
  can be tracked back to an (in retrospective somewhat unfortunate)
  choice of convention in \cite{tft1}. This is discussed in more
  detail in \cite[ch.\,3.1]{tft4}.}
\be
  \raisebox{-25pt}{
  \begin{picture}(102,58)
   \put(0,0){\scalebox{.75}{\includegraphics{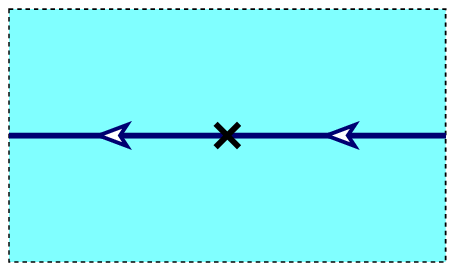}}}
   \put(0,0){
     \setlength{\unitlength}{.75pt}\put(-34,-15){
     \put( 45, 56) {\scriptsize$ a $}
     \put(142, 56) {\scriptsize$ b $}
     \put( 88, 36) {\scriptsize$ \phi^{a \leftarrow b} $}
     }\setlength{\unitlength}{1pt}}
  \end{picture}}   
  \quad \longmapsto \quad \eta^{ab} ~\cdot~
  \raisebox{-45pt}{
  \begin{picture}(160,90)
   \put(0,0){\scalebox{.75}{\includegraphics{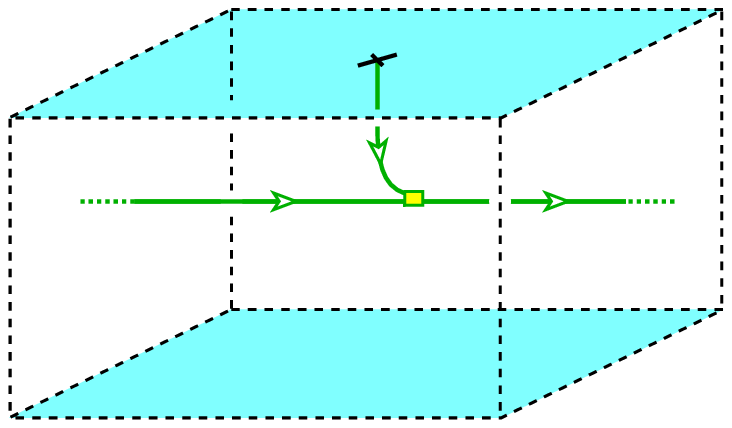}}}
   \put(0,0){
     \setlength{\unitlength}{.75pt}\put(-20,-18){
     \put( 72, 84) {\scriptsize$ a $}
     \put(183, 84) {\scriptsize$ b $}
     \put(115, 94) {\scriptsize$ f $}
     }\setlength{\unitlength}{1pt}}
  \end{picture}}
\ee
where $\eta^{ab} \in \Cb$ describes the normalisation of $\phi^{a\leftarrow b}$. The coupling to the identity in the OPE of two defect fields quoted in \erf{eq:def-norm} is then obtained by taking the summand for $k=0$ in
\bea
  \raisebox{-25pt}{
  \begin{picture}(102,58)
   \put(0,0){\scalebox{.75}{\includegraphics{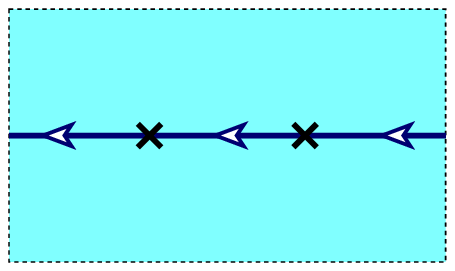}}}
   \put(0,0){
     \setlength{\unitlength}{.75pt}\put(-34,-15){
     \put( 45, 58) {\scriptsize$ a $}
     \put( 95, 58) {\scriptsize$ b $}
     \put(142, 58) {\scriptsize$ a $}
     \put( 67, 36) {\scriptsize$ \phi^{a \leftarrow b} $}
     \put(110, 36) {\scriptsize$ \phi^{b \leftarrow a} $}
     }\setlength{\unitlength}{1pt}}
  \end{picture}}
  \quad \longmapsto \quad 
\\[3em] \displaystyle
  \eta^{ab}\,\eta^{ba} ~
  \raisebox{-30pt}{
  \begin{picture}(160,90)
   \put(0,0){\scalebox{.75}{\includegraphics{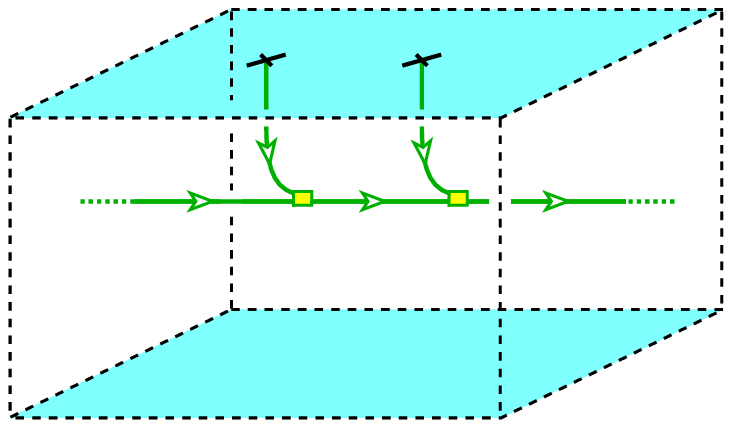}}}
   \put(0,0){
     \setlength{\unitlength}{.75pt}\put(-20,-18){
     \put( 61, 73) {\scriptsize$ a $}
     \put(129, 70) {\scriptsize$ b $}
     \put(182, 72) {\scriptsize$ a $}
     \put( 99,113) {\scriptsize$ f $}
     \put(143,113) {\scriptsize$ f $}
     }\setlength{\unitlength}{1pt}}
  \end{picture}}
  = \sum_{k \in \Ic} \eta^{ab}\,\eta^{ba}\,
  \Fs^{(ffa)a}_{bk} \,
  \raisebox{-30pt}{
  \begin{picture}(160,90)
   \put(0,0){\scalebox{.75}{\includegraphics{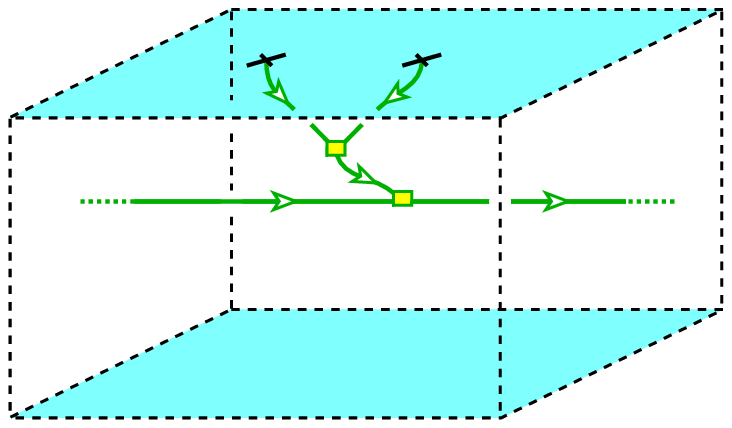}}}
   \put(0,0){
     \setlength{\unitlength}{.75pt}\put(-20,-18){
     \put( 69, 73) {\scriptsize$ a $}
     \put(182, 72) {\scriptsize$ a $}
     \put(124, 92) {\scriptsize$ k $}
     \put(101,113) {\scriptsize$ f $}
     \put(125,113) {\scriptsize$ f $}
     }\setlength{\unitlength}{1pt}}
  \end{picture}}
\nonumber
\eear\ee
Finally, the identities \erf{eq:collapse-field-bubble} amount to the following computations. (The surfaces are rotated by $180^\circ$ with respect to \erf{eq:collapse-field-bubble}. The orientation conventions is such that the surface gets embedded in the three-manifold `upside down', see \cite[ch.\,3.1]{tft4}.)
\be
  \raisebox{-25pt}{
  \begin{picture}(102,58)
   \put(0,0){\scalebox{.75}{\includegraphics{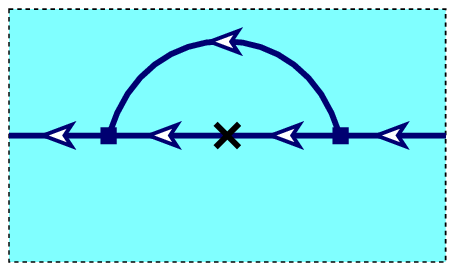}}}
   \put(0,0){
     \setlength{\unitlength}{.75pt}\put(-34,-15){
     \put( 45, 58) {\scriptsize$ c $}
     \put(142, 58) {\scriptsize$ d $}
     \put( 79, 58) {\scriptsize$ a $}
     \put(114, 58) {\scriptsize$ b $}
     \put(117, 76) {\scriptsize$ e $}
     \put( 88, 36) {\scriptsize$ \phi^{a \leftarrow b} $}
     }\setlength{\unitlength}{1pt}}
  \end{picture}}
  \quad \longmapsto \quad 
  \rule{20em}{0pt} 
  \nonumber
\ee  
\be
  \eta^{ab} ~
  \raisebox{-30pt}{
  \begin{picture}(160,90)
   \put(0,0){\scalebox{.75}{\includegraphics{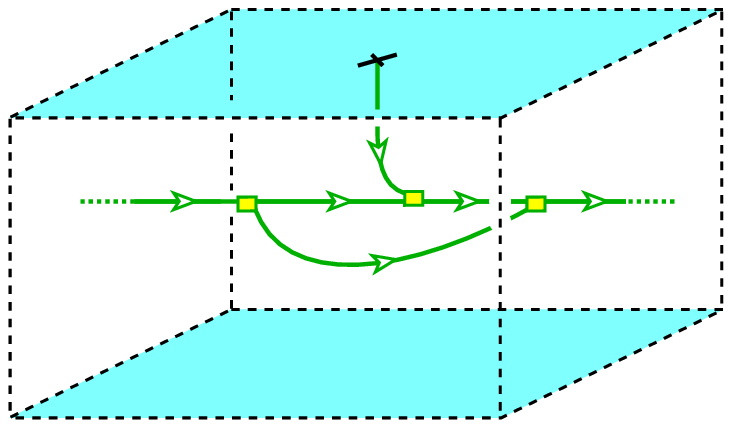}}}
   \put(0,0){
     \setlength{\unitlength}{.75pt}\put(-20,-18){
     \put(151, 86) {\scriptsize$ b $}
     \put(103, 86) {\scriptsize$ a $}
     \put(132, 57) {\scriptsize$ e $}
     \put( 72, 86) {\scriptsize$ c $}
     \put(186, 86) {\scriptsize$ d $}
     \put(115, 94) {\scriptsize$ f $}
     }\setlength{\unitlength}{1pt}}
  \end{picture}}
  = \eta^{ab} \, \Gs^{(fae)d}_{bc} ~
  \raisebox{-30pt}{
  \begin{picture}(160,90)
   \put(0,0){\scalebox{.75}{\includegraphics{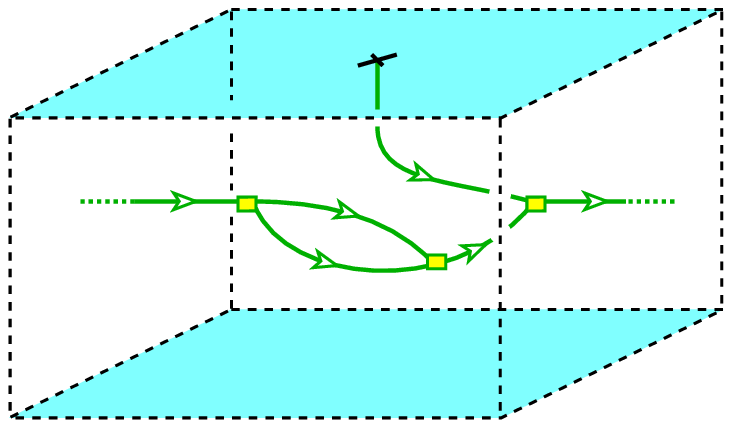}}}
   \put(0,0){
     \setlength{\unitlength}{.75pt}\put(-20,-18){
     \put(151, 72) {\scriptsize$ c $}
     \put(121, 80) {\scriptsize$ a $}
     \put(111, 55) {\scriptsize$ e $}
     \put( 72, 84) {\scriptsize$ c $}
     \put(186, 86) {\scriptsize$ d $}
     \put(118, 94) {\scriptsize$ f $}
     }\setlength{\unitlength}{1pt}}
  \end{picture}}
\nonumber
\ee
\be
  = \eta^{ab} \, \Gs^{(fae)d}_{bc} ~
  \raisebox{-30pt}{
  \begin{picture}(160,90)
   \put(0,0){\scalebox{.75}{\includegraphics{pic09b.eps}}}
   \put(0,0){
     \setlength{\unitlength}{.75pt}\put(-20,-18){
     \put( 72, 84) {\scriptsize$ c $}
     \put(183, 84) {\scriptsize$ d $}
     \put(115, 94) {\scriptsize$ f $}
     }\setlength{\unitlength}{1pt}}
  \end{picture}}
  \quad \longmapsto \quad 
  \frac{\eta^{ab}}{\eta^{cd}} \, \Gs^{(fae)d}_{bc} ~
  \raisebox{-25pt}{
  \begin{picture}(102,58)
   \put(0,0){\scalebox{.75}{\includegraphics{pic09a.eps}}}
   \put(0,0){
     \setlength{\unitlength}{.75pt}\put(-34,-15){
     \put( 45, 56) {\scriptsize$ c $}
     \put(142, 56) {\scriptsize$ d $}
     \put( 88, 36) {\scriptsize$ \phi^{c \leftarrow d} $}
     }\setlength{\unitlength}{1pt}}
  \end{picture}}   
\nonumber
\ee
and
\be
  \raisebox{-25pt}{
  \begin{picture}(102,58)
   \put(0,0){\scalebox{.75}{\includegraphics{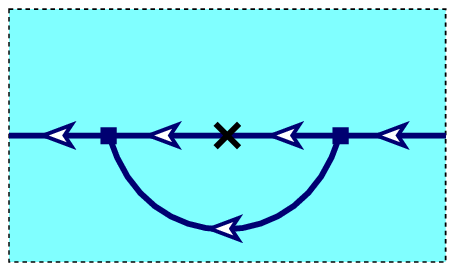}}}
   \put(0,0){
     \setlength{\unitlength}{.75pt}\put(-34,-15){
     \put( 45, 58) {\scriptsize$ c $}
     \put(142, 58) {\scriptsize$ d $}
     \put( 79, 58) {\scriptsize$ a $}
     \put(114, 58) {\scriptsize$ b $}
     \put(117, 23) {\scriptsize$ e $}
     \put( 88, 37) {\scriptsize$ \phi^{a \leftarrow b} $}
     }\setlength{\unitlength}{1pt}}
  \end{picture}}
  \quad \longmapsto \quad 
  \eta^{ab} ~
  \raisebox{-30pt}{
  \begin{picture}(270,90)
   \put(0,0){\scalebox{.75}{\includegraphics{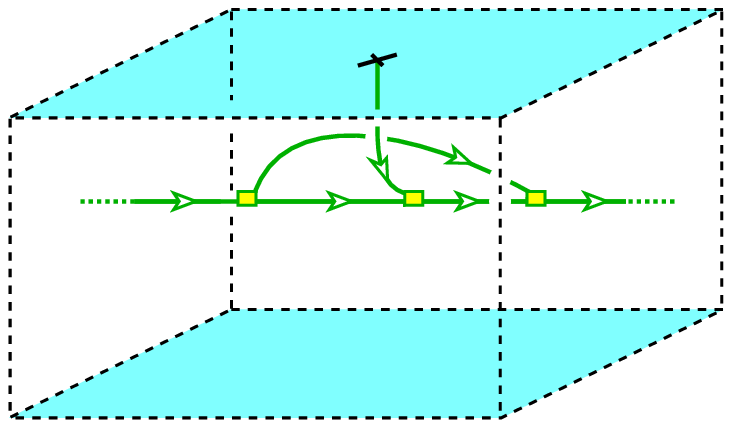}}}
   \put(0,0){
     \setlength{\unitlength}{.75pt}\put(-20,-18){
     \put(151, 70) {\scriptsize$ b $}
     \put(103, 72) {\scriptsize$ a $}
     \put(151, 97) {\scriptsize$ e $}
     \put( 72, 72) {\scriptsize$ c $}
     \put(186, 70) {\scriptsize$ d $}
     \put(129,111) {\scriptsize$ f $}
     }\setlength{\unitlength}{1pt}}
  \end{picture}}
\nonumber
\ee

\be
  =~ \eta^{ab} ~
  \raisebox{-30pt}{
  \begin{picture}(160,90)
   \put(0,0){\scalebox{.75}{\includegraphics{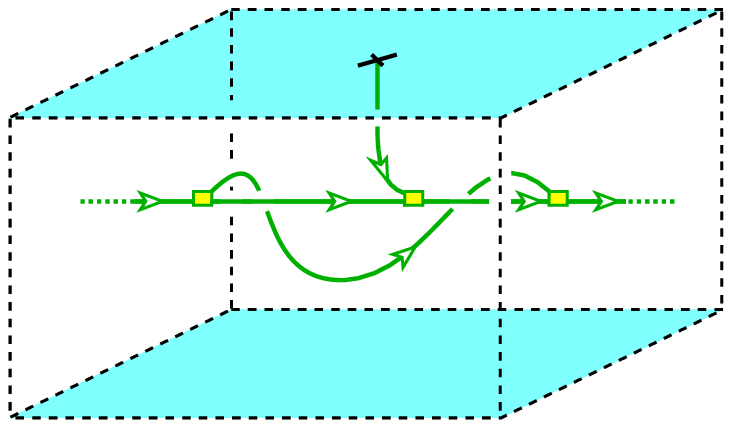}}}
   \put(0,0){
     \setlength{\unitlength}{.75pt}\put(-20,-18){
     \put( 62, 72) {\scriptsize$ c $}
     \put(190, 70) {\scriptsize$ d $}
     \put(168, 69) {\scriptsize$ b $}
     \put(103, 85) {\scriptsize$ a $}
     \put(140, 63) {\scriptsize$ e $}
     \put(115, 94) {\scriptsize$ f $}
     }\setlength{\unitlength}{1pt}}
  \end{picture}}
  ~=~ \eta^{ab} \, \frac{\Rs^{(be)d}}{\Rs^{(ae)c}}~
  \raisebox{-30pt}{
  \begin{picture}(300,90)
   \put(0,0){\scalebox{.75}{\includegraphics{pic11b.eps}}}
   \put(0,0){
     \setlength{\unitlength}{.75pt}\put(-20,-18){
     \put(151, 86) {\scriptsize$ b $}
     \put(103, 86) {\scriptsize$ a $}
     \put(132, 57) {\scriptsize$ e $}
     \put( 72, 86) {\scriptsize$ c $}
     \put(186, 86) {\scriptsize$ d $}
     \put(115, 94) {\scriptsize$ f $}
     }\setlength{\unitlength}{1pt}}
  \end{picture}}
\nonumber
\ee
  
\bea
  =~ \eta^{ab} \, \frac{\Rs^{(be)d}}{\Rs^{(ae)c}}
  \, \Gs^{(fae)d}_{bc} ~
  \raisebox{-30pt}{
  \begin{picture}(160,90)
   \put(0,0){\scalebox{.75}{\includegraphics{pic09b.eps}}}
   \put(0,0){
     \setlength{\unitlength}{.75pt}\put(-20,-18){
     \put( 72, 84) {\scriptsize$ c $}
     \put(183, 84) {\scriptsize$ d $}
     \put(115, 94) {\scriptsize$ f $}
     }\setlength{\unitlength}{1pt}}
  \end{picture}}
\\\displaystyle
  \hspace*{18em} \longmapsto ~~ 
  \frac{\eta^{ab}}{\eta^{cd}} \, \frac{\Rs^{(be)d}}{\Rs^{(ae)c}}
  \, \Gs^{(fae)d}_{bc} ~
  \raisebox{-25pt}{
  \begin{picture}(102,58)
   \put(0,0){\scalebox{.75}{\includegraphics{pic09a.eps}}}
   \put(0,0){
     \setlength{\unitlength}{.75pt}\put(-34,-15){
     \put( 45, 56) {\scriptsize$ c $}
     \put(142, 56) {\scriptsize$ d $}
     \put( 88, 36) {\scriptsize$ \phi^{c \leftarrow d} $}
     }\setlength{\unitlength}{1pt}}
  \end{picture}}   
\nonumber
\eear\ee

\subsection{F-matrix elements for minimal models}
\label{app:mm-chiral}

Let $a,b,\dots$ be entries in the Kac-table, $h_a, h_b,\dots$ be the corresponding conformal weights and $d_a, d_b,\dots$ the $d$-values as in \erf{eq:hrs-drs}. Fix also $1 \equiv (1,1)$, $2\equiv(1,2)$ and $3 \equiv (1,3)$. The braiding matrix is simply given by
\be
  \Rs^{(ab)c} = e^{\pi i (h_a+h_b-h_c)} ~.
\labl{eq:mm-Rs-val}
If $a = (r,s)$, then for $\eps= \pm 1$ denote by $a{+}\eps$ the Kac-label $(r,s{+}\eps)$. For $\eps,\nu = \pm 1$ one has
\be
  \Fs^{(2ac)b}_{b+\eps,a+\nu} =
  \Fs^{(a2b)c}_{b+\eps,a+\nu} =
  \Fs^{(cb2)a}_{b+\eps,a+\nu} =
  \Fs^{(bca)2}_{b+\eps,a+\nu}
  = \frac{ \Ga{\nu d_a} \, \Ga{1{-}\eps d_b} }{
    \Ga{\tfrac12(1{+}d_c{+}\nu d_a{-}\eps d_b)}
    \Ga{\tfrac12(1{-}d_c{+}\nu d_a{-}\eps d_b)} }
    ~.
\labl{eq:mm-F-2}
This follows as usual from the transformation behaviour of a basis of solutions to a level 2 null-vector equation \cite[ch.\,8]{DiFr}. From these basic $\Fs$-matrix entries all others can be obtained recursively via the pentagon identity, see e.g.\ \cite{Runkel:1998pm}. The relation between the notation $\Fs$ used here and that of \cite{Moore:1989vd} (and also \cite{Runkel:1998pm}) is
\be
   \Fs^{(ijk)l}_{pq} = 
   \Fs_{pq}\Big[\begin{array}{cc} i &j \\ l &k \end{array}\Big]
   ~~.
\ee
Rather than using a recursive procedure, one can also directly compute with the closed form expression for the $\Fs$-matrices \cite{Dotsenko:1984ad,Furlan:1989ra} (collected e.g.\ in \cite[app.\,A.1.1]{Graham:2001tg}), but for the present application the recursive procedure is more convenient. The inverse $\Gs$ of the $\Fs$-matrix is related to $\Fs$ in a simple way (combine \cite[eqn.\,(2.61)]{tft1} with \erf{eq:mm-Rs-val})
\be
  \Gs^{(ijk)l}_{pq} = \Fs^{(kji)l}_{pq}  ~.
\labl{eq:mm-GF}
Some specific $\Fs$-matrix entries used in the main text are, for $a \equiv (r,s)$,
\be
  \Fs^{(33a)a}_{a1} = 
  \frac1{1{-}3t}\,
  \frac{
  \Ga{1{-}t{+}d_a}\, \Ga{1{-}t{-}d_a} \,\Ga{1{-}t} \,\Ga{2t}\, \Ga{3t}
  }{
  \Ga{t{-}d_a}\, \Ga{t{+}d_a}\, \Ga{2{-}2t}\, \Ga{t}\, \Ga{t}
  } ~~,
\labl{eq:F(33x)}
as well as, for $r \equiv (r,1)$, $s \equiv (1,s)$ and $a \equiv (r,s)$,
\be
  \Fs^{(rs3)a}_{sa} = 
  \frac{
  \Ga{1{-}t{+}d_{1,s}}\, \Ga{t{+}d_{r,s}}
  }{
  \Ga{1{-}t{+}d_{r,s}}\, \Ga{t{+}d_{1,s}}
  } ~~.
\labl{eq:F(rs3)}
These have been obtained by starting from the values \erf{eq:mm-F-2} and applying one step in the recursive procedure mentioned above. Finally, we also need, for $p'{-}1 \equiv (1,p'{-}1)$, $s \equiv (1,s)$ and $p'{-}s \equiv (1,p'{-}s)$,
\be
  \Fs^{(p'-1,s,3)p'-s}_{s,p'-s}
  = (-1)^p \,
  \frac{ \Ga{2{-}t{-}st} \, \Ga{st{-}t} }{
  \Ga{2{-}t{-}(p'{-}s)t} \, \Ga{(p'{-}s)t{-}t} } ~.
\labl{eq:F(ps3)}

\end{document}